\documentclass[11pt]{article}
\usepackage{graphicx}
\usepackage{mathtools}
\usepackage{extsizes}
\usepackage{geometry}
\usepackage{amsmath}

\usepackage{amssymb}
\usepackage[utf8]{inputenc}
\usepackage{cite}
\usepackage{times}
\usepackage{setspace}
\usepackage{caption}
\usepackage{fancyhdr}
\usepackage{authblk}
\usepackage{xcolor}

\date{}
 
\usepackage{float}
\usepackage[colorlinks=true,citecolor=blue,linkcolor=blue,urlcolor=blue]{hyperref}

\title{\textbf{Matter-Only Late-Time Acceleration via Barrow Entropy in Kaniadakis Statistics}}

\author[1]{{Navaneeth Preman P K}}
\author[2]{Sarath Nelleri}
\author[3]{Sreerag Radhakrishnan}
\author[4]{Pranav Prasanthan}
\author[5]{Nandhida Krishnan P}
\author[1]{Emmanuel Tom}

\affil[1]{Department of Physics, Nirmalagiri college, Kuthuparamba-670701, India}
\affil[2]{Turing Enterprises, USA (Remote based in India)}
\affil[3]{Department of Earth and Space Sciences, Indian Institute of Space Science and Technology, Thiruvananthapuram-695541, India}
\affil[4]{Institute of Physics, University of Szczecin, Wielkopolska 15, Szczecin,  70-451, Poland}
\affil[5]{Department of Physics, Cochin University of Science and Technology, Kochi 682-022, India}

\date{}

\begin{document}

\maketitle

\begin{abstract}

The Barrow correction to the horizon entropy changes the total number of microstates associated with the horizon, whereas the Kaniadakis correction provides an alternate statistical description of these microstates different from Boltzmann-Gibbs statistics. In this paper, we consider the Kaniadakis description of the Barrow-corrected horizon entropy analogous to its description of the standard horizon entropy. The modified Friedmann equations were derived for an $(n+1)$-dimensional non-flat universe through two different approaches --- the law of emergence and the first law of thermodynamics. The generalised second law of thermodynamics was investigated, and the density contrast evolution equation was also derived. We take the universe to be composed of only matter and constrain the model parameters through an MCMC analysis using observational Hubble data, Pantheon+ Type Ia supernovae data, baryon acoustic oscillation distances from the DESI DR2 data release, and $f\sigma_{8}$ measurements from redshift-space distortions. The Kaniadakis parameter $K$ was constrained away from zero, showing a preference for the $K$-deformed statistical approach. Motivated by recent studies, we considered a wider parameter space for the Barrow index, $\Delta \in (-1,1]$. In agreement with observational data, we obtained $\Delta = -0.335\pm0.031$, motivating a horizon structure different from Barrow's sphereflake configuration. The present value of the Hubble parameter was constrained as $H_{0}=(66.98 \pm 0.71)$ km/s/Mpc, which leans towards the early-universe Planck estimates; however, late-time data alone are insufficient to study the Hubble tension problem. Analysing the evolution of various cosmological parameters, we conclude that the modified Friedmann dynamics of the model produces a viable expansion history and late-time acceleration without invoking a dark energy component.

\end{abstract}

\section{Introduction} \label{s1}

Bekenstein \cite{Bekenstein:1973ur} proposed that black holes possess an entropy proportional to the area of the event horizon, while Hawking \cite{Hawking:1975vcx} showed that black holes emit thermal radiation at a temperature proportional to the surface gravity. These findings established a connection between gravity and thermodynamics, which was further explored by Jacobson \cite{Jacobson:1995ab}, who derived the Einstein field equations by applying the Clausius relation, $dQ=TdS$, to local Rindler horizons. This showed that the Einstein field equations can be interpreted as a thermodynamic equation of state. Following these developments, Thanu Padmanabhan \cite{Padmanabhan:2009kr, Padmanabhan:2012ik} proposed that cosmic expansion results from the emergence of cosmic space, and that the evolution of the universe is driven by the difference between bulk degrees of freedom and the degrees of freedom residing on the horizon surface, $dV/dt \propto (N_{surf}-N_{bulk})$, called the law of emergence. In this framework, the de Sitter universe represents a state of holographic equipartition. The attainment of holographic equipartition has also been associated with the maximisation of horizon entropy \cite{KrishnaPB:2020sne}. Since the surface degrees of freedom are determined by the horizon entropy, modifying the entropy-area relation allows the law of emergence to yield modified Friedmann equations and consequently modified cosmic dynamics \cite{Komatsu:2016vof, Sheykhi:2018dpn, Sheykhi:2021fwh, Luciano:2023zrx, P:2023jeq, Lymperis:2021qty, Prasanthan:2024xsl}. 

The area scaling of Bekenstein--Hawking entropy ($S_{BH}=A/4L_{p}^{2}$) \cite{Bekenstein:1973ur, Hawking:1975vcx}, together with the long-range nature of gravitational interactions, motivates the consideration of non-extensive statistical frameworks, beyond the standard Boltzmann–Gibbs formalism \cite{Tsallis:1987eu, Tsallis:2012js, sharma1975new, Ghaffari_2019, Kaniadakis:2001tcv, Kaniadakis:2002zz, Moradpour:2020dfm}. The Kaniadakis statistics arises from a coherent and self-consistent relativistic statistical theory and provides a generalised form of entropy well-suited for relativistic systems. The modified horizon entropy within Kaniadakis statistics is given by
 \begin{equation}
     S_{K} = \dfrac{1}{K}\sinh\left\{KS_{BH}\right\}, \label{1.01}
 \end{equation}
where $-1<K<1$ is the Kaniadakis parameter. Modified Friedmann equations have been derived in the context of Kaniadakis entropy using both the first law of thermodynamics \cite{Lymperis:2021qty, Sheykhi:2023aqa} and the law of emergence \cite{Prasanthan:2024xsl, Prasanthan:2024fwg}. Observational constraints on Kaniadakis-based cosmological models have been investigated \cite{Hernandez-Almada:2021rjs, Hernandez-Almada:2021aiw, Zarandi:2025ytt}, and the Generalised Second Law (GSL) of thermodynamics has also been studied \cite{Sheykhi:2023aqa, Prasanthan:2024xsl, Odintsov:2024hzu, Prasanthan:2024fwg}.
 
Apart from the modifications based on generalised statistical frameworks, John D. Barrow \cite{Barrow:2020tzx} proposed a modification of horizon entropy by considering possible deformations of the smooth black hole horizon due to quantum effects near Planck scales, through his "sphereflake" model. The corresponding modified horizon entropy is given by
 \begin{equation}
     S_{B}=\left(\frac{A}{4L_{p}^{2}}\right)^{1+\frac{\Delta}{2}}, \label{1.02}
 \end{equation}
where $L_{p}$ is the Planck length and the index $\Delta$, called the Barrow index, was originally introduced with the bounds $0\leq\Delta\leq1$ and measures the fractional excess to the horizon area. Modified Friedmann equations were derived for Barrow entropy through both the first law of thermodynamics \cite{Saridakis:2020lrg, Sheykhi:2021fwh} and the law of emergence \cite{Luciano:2023zrx, P:2023jeq}. Observational constraints on Barrow-entropy based cosmological models have also been studied \cite{Asghari:2021bqa, Leon:2021wyx, Luciano:2025hjn, Anagnostopoulos:2020ctz}. The sphereflake construction restricted $\Delta$ to positive values, but more recently, Refs.~\cite{Luciano:2025hjn, Luciano:2026vhm, Jizba:2024klq, Bolotin:2026hot} have explored negative values for the Barrow index, suggesting a horizon surface characterised by internal voids or porosity rather than a sphereflake-type configuration. Furthermore, real-world porous materials having a Hausdorff dimension close to $1$ (i.e., $\Delta \approx -1$) \cite{TANG2012383, article} and support from QFT where renormalization group arguments allow for a negative $\Delta$ for certain systems \cite{DAGOTTO1990268}, motivate considering a broader parameter space, $\Delta \in (-1, 1]$. The GSL of thermodynamics has also been investigated for Barrow entropy \cite{Saridakis:2020cqq, Sheykhi:2021fwh}. It was shown in Ref.~\cite{Saridakis:2020cqq} that the GSL remains valid in the case of Barrow entropy within a $\Lambda$CDM background but can be violated for certain cosmological evolutions.

Recently, Kolesnichenko \cite{AR308339} proposed a horizon entropy combining Barrow and Kaniadakis corrections by replacing the Bekenstein-Hawking entropy with the Barrow entropy in the dual Kaniadakis entropy expression, invoking entropic force to drive late-time accelerated expansion. Motivated by this study, we investigate late-time accelerated expansion in a matter-only universe through modified Friedmann dynamics arising from horizon entropy modifications involving both Barrow and standard Kaniadakis corrections within the emergent gravity paradigm. We derive the Friedmann equations for an $(n+1)$-dimensional non-flat universe using both the law of emergence and the first law of thermodynamics. The generalised second law of thermodynamics is investigated, and the evolution equation for linear matter density perturbations is also derived. The model parameters are constrained using a joint analysis of non-homogeneous observational Hubble data (OHD), Pantheon+ Type Ia supernovae (SNe Ia) data, baryon acoustic oscillation (BAO) measurements from DESI DR2, and $f\sigma_{8}$ measurements from large-scale structure (LSS) observations using redshift-space distortions (RSD). The evolution of various cosmological parameters is also studied.

The paper is organised as follows. Section \ref{s2} introduces the Barrow-modified horizon entropy within Kaniadakis statistics. Sections \ref{s3} and \ref{s4} derive the modified Friedmann equations using the law of emergence and the first law of thermodynamics, respectively. Section \ref{s5} investigates the generalised second law of thermodynamics, while Section \ref{s6} derives the evolution equation for linear matter density perturbations. Section \ref{s7} presents the observational constraints on the model parameters, and Section \ref{s8} examines the cosmological evolution of the model, including the expansion history, growth of matter perturbations, and statefinder diagnostic. Finally, Section \ref{s9} discusses the results and presents the conclusions.

\section{Modifications to Horizon Entropy} \label{s2}

To account for the deformation of the horizon surface due to quantum effects, we consider the Barrow modification of the horizon entropy \cite{Barrow:2020tzx}. In an $(n+1)$-dimensional universe, the Barrow entropy given in Eq.~\eqref{1.02} takes the form \cite{Luciano:2023roh}
\begin{equation}
    S_{B}=\left(\dfrac{A_{n+1}}{4L_{p}^{n-1}}\right)^{1+\frac{\Delta}{2}}, \label{2.01}
\end{equation}
where $A_{n+1}=n\Omega_{n}r_{a}^{n-1}$ denotes the area of the apparent horizon, with the radius for a non-flat FLRW universe given by \cite{Hayward:1998ee,Hayward:1993wb} \begin{equation}\label{app}
    r_{a}=\frac{1}{\sqrt{H^2 + \frac{k}{a^2}}}.
\end{equation}
To connect this geometric description with statistical mechanics, we relate the Barrow-modified horizon entropy to the Gibbs–Shannon entropy of the associated microstates:
\begin{equation}
    S_{B}=-\sum_{i}n_{i}\ln n_{i}, \label{2.02}
\end{equation}
where $n_{i}$ is the probability that a system is in a specific microstate. Since we are interested in late-time analysis, we can consider quasi-static expansion and therefore quasi-static equilibrium for applying equal a priori probabilities for the microstates associated with the horizon. Therefore, using Eq.~\eqref{2.02} we can obtain $n_{i}=e^{-S_{B}}$ \cite{Lymperis:2021qty, AR308339}. We now incorporate the Kaniadakis description of the horizon entropy with Barrow entropy \eqref{2.01}. The $K$-deformed exponential and $K$-deformed logarithm are \cite{Kaniadakis:2001tcv}
\begin{equation}
    e_{\{K\}}(x) = \left(\sqrt{1+K^{2}x^{2}}+Kx\right)^{\frac{1}{K}},
    \quad
    \ln_{\{K\}}(x) = \dfrac{x^{K}-x^{-K}}{2K}.
    \label{2.06}
\end{equation}
These functions reduce to the standard exponential and logarithm as the real deformation parameter $K$ approaches zero. The Kaniadakis entropy is given by \cite{Kaniadakis:2002zz}
\begin{equation}
    S_{K}=-\sum_{i}n_{i}\ln_{\left\{K\right\}}n_{i}. \label{2.08}
\end{equation}
Substituting the Kaniadakis logarithm from Eq.~\eqref{2.06} into Eq.~\eqref{2.08}, with $n_{i}=e^{-S_{B}}$, we obtain both the Kaniadakis and Barrow corrections to the horizon entropy as
\begin{equation}
    S_{KB}=-\sum_{i}\dfrac{\left(e^{-S_{B}}\right)^{K+1}-\left(e^{-S_{B}}\right)^{1-K}}{2K}, \label{2.09}
\end{equation}
which can be further simplified \cite{Moradpour:2020dfm, Lymperis:2021qty} to arrive at the following result 
\begin{equation}
    S_{KB}=\dfrac{1}{K}\sinh\left(K \left(\dfrac{A_{n+1}}{4L_{p}^{n-1}}\right)^{1+\frac{\Delta}{2}}\right). \label{2.11}
\end{equation}
This entropy expression incorporates both the Barrow and Kaniadakis modifications. The Barrow index $\Delta$ characterises the deviation of the horizon geometry from a smooth surface, while the Kaniadakis parameter $K$ characterises the deformation of Boltzmann--Gibbs statistics. In the following sections, we study the cosmological dynamics resulting from this modified horizon entropy.

\section {Friedmann equation from the Law of Emergence} \label{s3}

The law of emergence, proposed by Padmanabhan \cite{Padmanabhan:2012ik}, relates the expansion of the universe to the difference between the degrees of freedom on the cosmic horizon ($N_{surf}$) and those in the bulk ($N_{bulk}$):
\begin{equation}
    \frac{dV}{dt}=L_{p}^{2}\left(N_{surf}-N_{bulk}\right). \label{3.00}
\end{equation}
For an $(n+1)$-dimensional non-flat universe, this law was extended by taking the apparent horizon as the boundary of the thermodynamic system \cite{Cai:2012ip, Sheykhi:2013oac}, giving
\begin{equation}
    \alpha\frac{d\tilde{V}}{dt} = L_{p}^{n-1}\frac{r_{a}}{H^{-1}} \left(N_{surf} - N_{bulk}\right), \label{3.01}
\end{equation}
where $\alpha=(n-1)/[2(n-2)]$ is a dimension-dependent normalisation factor. Following Refs.~\cite{Cai:2012ip, Sheykhi:2021fwh, Prasanthan:2024xsl}, we associate the modified entropy $S_{KB}$ in Eq.~\eqref{2.11} with an effective horizon area $\tilde A$, defined by $S_{KB}=\tilde A/(4L_p^{n-1})$. This gives
\begin{equation}
    \tilde{A} = \frac{4L_{p}^{n-1}}{K} \sinh\left[K\left(\frac{n\Omega_{n}r_{a}^{n-1}}{4L_{p}^{n-1}}\right)^{1+\frac{\Delta}{2}}\right]. \label{3.03}
\end{equation}
The corresponding effective volume $\tilde V$ is defined through
\begin{equation}
    \frac{d\tilde V}{dt}=\frac{r_{a}}{n-1}\frac{d\tilde A}{dt}.
    \label{3.04}
\end{equation}
For convenience, in the following expressions, we denote the argument of the hyperbolic functions by
\begin{equation}
    X(r_{a})\equiv K\left(\frac{n\Omega_{n} r_{a}^{n-1}}{4L_{p}^{n-1}}\right)^{1+\frac{\Delta}{2}}.
\end{equation}
Substituting Eq.~\eqref{3.03} into Eq.~\eqref{3.04}, from the resulting expression we find
\begin{equation}
\begin{split}
    \frac{d\tilde{V}}{dt} =&~~ r_{a}^{n+2} \left(\dfrac{\left(n\Omega_{n}\right)^{1+\frac{\Delta}{2}}}{\left(n-1\right)\frac{\Delta}{2}-2}\right) \left(\dfrac{1+\frac{\Delta}{2}}{\left(4L_{p}^{n-1}\right)^{\frac{\Delta}{2}}}\right)\\ 
    &\times \dfrac{d}{dt}\left[\dfrac{\cosh\left\{X(r_{a})\right\}}{r_{a}^{2}r_{a}^{\left(n-1\right)\frac{\Delta}{2}}} - \int_{0}^{r_{a}}\dfrac{\sinh\left\{X(x)\right\}}{x^{2}x^{\left(n-1\right)\frac{\Delta}{2}}}d\left\{X(x)\right\}\right], \label{3.06}
\end{split}
\end{equation}
where $x$ is the variable of integration. From Eq.~\eqref{3.06}, following Refs.~\cite{Sheykhi:2013oac, Prasanthan:2024xsl, P:2023jeq}, we propose the surface degrees of freedom associated with the apparent horizon as
\begin{equation}
\begin{split}
    N_{surf} =& \left(\frac{n-1}{n-2}\right)\left(\frac{1+\frac{\Delta}{2}}{2-\left(n-1\right)\frac{\Delta}{2}}\right)\left(\dfrac{n\Omega_{n}}{L_{p}^{n-1}}\right)\left(\frac{n\Omega_{n}}{4L_{p}^{n-1}}\right)^{\frac{\Delta}{2}}\\
    &\times \left[\frac{\cosh\left\{X(r_{a})\right\}}{r_{a}^{-\left(n-1\right)\left(1+\frac{\Delta}{2}\right)}} - r_{a}^{n+1}\int_{0}^{r_{a}}\frac{\sinh\left\{X(x)\right\}}{x^{2}x^{-\left(n-1\right)\frac{\Delta}{2}}}d\left\{X(x)\right\}\right]. \label{3.07}
\end{split}
\end{equation}
The bulk degrees of freedom for an $(n+1)$-dimensional non-flat universe is given by \cite{Cai:2012ip, Sheykhi:2013oac}
\begin{equation}
N_{bulk} = -\frac{4\pi\Omega_{n}}{n-2}r_{a}^{n+1}\left[(n-2)\rho+nP\right], \label{3.08}
\end{equation}
where the temperature associated with the apparent horizon is taken as the Cai-Kim temperature $T=1/(2\pi r_{a})$, which is the Hawking-like temperature associated with the apparent horizon \cite{Cai:2005ra, Cai:2008gw}. \footnote{For completeness, we note that recent studies have emphasised the importance of considering the underlying mass-to-horizon relation when employing generalised horizon entropies together with the Hawking
temperature, particularly to ensure compatibility with the Clausius relation; see
Refs.~\cite{Gohar:2023hnb, Gohar:2023lta}.} Substituting Eqs.~\eqref{3.07} and \eqref{3.08} into Eq.~\eqref{3.01}, and using the continuity equation $\dot{\rho} + nH\left(\rho + P\right) = 0$, we arrive at
\begin{equation}
\begin{split}
    &\frac{4\pi L_{p}^{n-1}}{n\left(n-1\right)}\left(\frac{\dot{\rho}a}{\dot{a}}+2\rho\right)= -\left(1+\frac{\Delta}{2}\right)\left(\frac{n\Omega_{n}}
    {4L_{p}^{n-1}}\right)^{\frac{\Delta}{2}}\frac{H^{-1}}{2}\left[\frac{\cosh\left\{X(r_{a})\right\}}{r_{a}^{3}r_{a}^{-\left(n-1\right)\left(\frac{\Delta}{2}\right)}}\right]\dot{r}_{a}\\
    &+ \left(\frac{1+\frac{\Delta}{2}}{2-\left(n-1\right)\frac{\Delta}{2}}\right)\left(\frac{n\Omega_{n}}
    {4L_{p}^{n-1}}\right)^{\frac{\Delta}{2}}\left[\frac{\cosh\left\{X(r_{a})\right\}}{r_{a}^{2}r_{a}^{-\left(n-1\right)\frac{\Delta}{2}}} - \int_{0}^{r_{a}}\frac{\sinh\left\{X(x)\right\}}{x^{2}x^{-\left(n-1\right)\frac{\Delta}{2}}}d\left\{X(x)\right\}\right]. \label{3.10}
\end{split}
\end{equation}
Multiply both sides by $2a\dot{a}$ and then add and subtract the following term on the RHS of Eq.~\eqref{3.10},
\begin{equation}
\begin{split}
    &a^{2}\left(\frac{1+\frac{\Delta}{2}}{2-\left(n-1\right)\frac{\Delta}{2}}\right)\left(\frac{n\Omega_{n}}
    {4L_{p}^{n-1}}\right)^{\frac{\Delta}{2}}\left[\frac{\sinh\left\{X(r_{a})\right\}}{r_{a}^{2}r_{a}^{-\left(n-1\right)\frac{\Delta}{2}}}\right] \frac{d\left\{X(r_{a})\right\}}{dt}. \label{3.11}
\end{split}
\end{equation}
Integrating the resulting expression and setting the integration constant to zero, we obtain the modified first Friedmann equation in the present context:
\begin{equation}
\begin{split}
    &\frac{8\pi L_{p}^{n-1}}{n\left(n-1\right)}\rho = \left(\frac{1+\frac{\Delta}{2}}{2-\left(n-1\right)\frac{\Delta}{2}}\right)\left(\frac{n\Omega_{n}}
    {4L_{p}^{n-1}}\right)^{\frac{\Delta}{2}} \left[\frac{\cosh\left\{X(r_{a})\right\}}{r_{a}^{2}r_{a}^{-\left(n-1\right)\frac{\Delta}{2}}} - \int_{0}^{r_{a}}\frac{\sinh\left\{X(x)\right\}}{x^{2}x^{-\left(n-1\right)\frac{\Delta}{2}}}d\left\{X(x)\right\}\right]. \label{3.13}
\end{split}
\end{equation}
For a $(3+1)$-dimensional flat universe, $n=3$ and $k=0$, so that $r_{a}=H^{-1}$ and $X=K\left(\pi/(H^{2}L_{p}^{2})\right)^{1+\Delta/2}$. Then Eq.~\eqref{3.13} reduces to the following result
\begin{equation}
\begin{split}
    &\dfrac{8\pi L_{p}^{2}}{3}\rho = \left(\dfrac{2+\Delta}{2-\Delta}\right)\left(\dfrac{\pi}{L_{p}^{2}}\right)^{\frac{\Delta}{2}} \left[\dfrac{\cosh\left\{X(H)\right\}}{H^{-2}H^{\Delta}} - \int_{0}^{H^{-1}}\dfrac{\sinh\left\{X(x)\right\}}{x^{-2}x^{\Delta}}d\left\{X(x)\right\}\right]. \label{3.14}
\end{split}    
\end{equation}
For $\Delta=0$, the Barrow correction vanishes, and the equation retains only the Kaniadakis modification. Taking also the limit $K\to0$, we recover the standard first Friedmann equation
\begin{equation}
    H^{2}=\frac{8\pi G}{3}\rho. \label{3.16}
\end{equation}

\section{Friedmann equation from the First Law of Thermodynamics} \label{s4}

It is well established that there is a connection between the law of emergence and the first law of thermodynamics \cite{Krishna:2022zzw, T:2022mqr}. To examine whether this connection also holds within the present framework, we derive the modified Friedmann equation using the first law of thermodynamics and compare it with Eq.~\eqref{3.13} obtained from the law of emergence. We adopt the Cai--Kim approach \cite{Cai:2005ra}, in which the Friedmann equations are derived by applying the Clausius relation to the apparent horizon of an $(n+1)$-dimensional non-flat FLRW universe. The heat transfer is evaluated over an infinitesimal time interval $dt$, during which the apparent horizon is treated as fixed. Under this condition, the Clausius relation takes the form \cite{Cai:2005ra}
\begin{equation}
    \delta Q = -dE = TdS,
    \label{4.01}
\end{equation}
where $T=1/(2\pi r_{a})$ is the Cai--Kim temperature. For the time interval $dt$, the energy change inside the horizon is entirely due to the flux of the cosmic
fluid across the apparent horizon. This is defined in terms of the time component of the energy flow vector \cite{Cai:2005ra, Tian:2014ila} as
\begin{equation}
    \delta Q = -dE = -A_{n+1}\psi_{t} = n\Omega_{n} r_{a}^{n} H(\rho+P)dt. \label{4.02}
\end{equation}
Using Eq.~\eqref{2.11}, we get
\begin{equation}
\begin{split}
&dS = \left(n-1\right)\left(1+\frac{\Delta}{2}\right)\left(\dfrac{n\Omega_{n}}{4L_{p}^{n-1}}\right)^{1+\frac{\Delta}{2}}\left[\dfrac{\cosh{\left\{X(r_{a})\right\}}}{r_{a}^{-\left(n-1\right)\left(1+\frac{\Delta}{2}\right)+1}}\right]dr_{a}. \label{4.08}
\end{split}
\end{equation}
Substituting Eq.~\eqref{4.08} into Eq.~\eqref{4.02} and using the continuity equation $\dot{\rho} + nH(\rho + P) = 0$, we obtain the following
\begin{equation}
\begin{split}
&\Omega_{n}r_{a}^{n}d\rho = -\dfrac{1}{2\pi r_{a}}\left(n-1\right)\left(1+\frac{\Delta}{2}\right)\left(\dfrac{n\Omega_{n}}{4L_{p}^{n-1}}\right)^{1+\frac{\Delta}{2}}\left[\frac{\cosh{\left\{X(r_{a})\right\}}}{r_{a}^{-\left(n-1\right)\left(1+\frac{\Delta}{2}\right)+1}}\right]dr_{a}. \label{4.11}
\end{split}
\end{equation}
After integrating and setting the integration constant to zero, we get the modified first Friedmann equation:
\begin{equation}
\begin{split}
    \frac{8\pi L_{p}^{n-1}}{n\left(n-1\right)}\rho = \left(\frac{1+\frac{\Delta}{2}}{2-\left(n-1\right)\frac{\Delta}{2}}\right)\left(\frac{n\Omega_{n}}
    {4L_{p}^{n-1}}\right)^{\frac{\Delta}{2}} \left[\frac{\cosh\left\{ X(r_{a})\right\}}{r_{a}^{2}r_{a}^{-\left(n-1\right)\frac{\Delta}{2}}} -\int_{0}^{r_{a}}\frac{\sinh\left\{ X(x)\right\}}{x^{2}x^{-\left(n-1\right)\frac{\Delta}{2}}}d\left\{ X(x)\right\}\right]. \label{4.12}
\end{split}
\end{equation}
This is the same result as obtained from the law of emergence. This suggests that Barrow entropy within Kaniadakis statistics is consistent with both the law of emergence and the first law of thermodynamics. From Eq.~\eqref{4.11}, using the continuity equation, we get
\begin{equation}
\begin{split}
    &-\dfrac{8\pi L_{p}^{n-1}}{n-1}\left(\rho + P\right) = \left(1+\frac{\Delta}{2}\right)\left(\frac{n\Omega_{n}}{4L_{p}^{n-1}}\right)^{\frac{\Delta}{2}}r_{a}^{\left(n-1\right)\frac{\Delta}{2}}\left(\dot{H}-\dfrac{k}{a^{2}}\right) \cosh\left\{X(r_{a})\right\}. \label{4.13}
\end{split}
\end{equation}
Substituting for $8 \pi L_{p}^{n-1}\rho/(n-1)$ from Eq.~\eqref{4.12} and re-arranging, we can arrive at the modified second Friedmann equation:
\begin{equation}
\begin{split}
&-\dfrac{8\pi L_{p}^{n-1}}{n-1}P = \left(\frac{1+\frac{\Delta}{2}}{2-\left(n-1\right)\frac{\Delta}{2}}\right)\left(\frac{n\Omega_{n}}{L_{p}^{n-1}}\right)^{\frac{\Delta}{2}}\\
&\times\left[\dfrac{\cosh\left\{X(r_{a})\right\}}{r_{a}^{-\left(n-1\right)\frac{\Delta}{2}}} \left( nr_{a}^{-2}+ \left(2-\left(n-1\right)\frac{\Delta}{2}\right)\left(\dfrac{\ddot{a}}{a}-H^{2}-\dfrac{k}{a^{2}}\right)\right) - n\int_{0}^{r_{a}}\dfrac{\sinh\left\{X(x)\right\}}{x^{2}{x^{-\left(n-1\right)\frac{\Delta}{2}}}}d\left\{X(x)\right\} \right]. \label{4.14}
\end{split}
\end{equation}
For a $(3+1)$-dimensional flat universe, Eq.~\eqref{4.14} takes the form
\begin{equation}
\begin{split}
    P = &\left[\dfrac{\cosh\left\{X(H)\right\}}{H^{\Delta}} \left( 3H^{2} + \left(2-\Delta\right)\left(\dfrac{\ddot{a}}{a}-H^{2}\right)\right) - 3\int_{0}^{H^{-1}}\dfrac{\sinh\left\{X(x)\right\}}{x^{-2}x^{\Delta}}d\left\{X(x)\right\}\right]\\ &\times \left(\dfrac{1+\frac{\Delta}{2}}{2-\Delta}\right)\left(\dfrac{\pi}{L_{p}^{2}}\right)^{\frac{\Delta}{2}}. \label{4.15}
\end{split}
\end{equation}
For $\Delta=0$ and $K\to0$, the Kaniadakis and Barrow corrections vanish. Substituting the standard first Friedmann equation, $H^2=8\pi G\rho/3$, into the resulting expression gives the standard second Friedmann equation
\begin{equation}
    \frac{\ddot{a}}{a} = -\frac{4\pi G}{3}(\rho + 3P). \label{4.18}
\end{equation}

\section{Generalised Second Law of Thermodynamics (GSL)} \label{s5}

The generalized second law of thermodynamics (GSL) requires that the total entropy of the horizon $(S_{h})$ and the matter enclosed within it $(S_{m})$ does not decrease with time \cite{Davies:1987ti, Bekenstein:1974ax}, that is,
\begin{equation}
    \dot{S}_{tot} = \dot{S}_{h} + \dot{S}_{m} \geq 0. \label{5.01}
\end{equation}
Following Refs.~\cite{Sheykhi:2021fwh, Prasanthan:2024xsl}, we examine whether the GSL is satisfied within the present framework. Using the continuity equation $\dot{\rho} + nH(\rho + P)=0$ in Eq.~\eqref{4.11}, we obtain the evolution of the apparent horizon radius as
\begin{equation}
\begin{split}
    \dot{r}_{a}= \frac{8\pi L_{p}^{n-1}}{n-1}\dfrac{r_{a}^{3-(n-1)\frac{\Delta}{2}}H(\rho+P)}{\left(1+\dfrac{\Delta}{2}\right)\left(\dfrac{n\Omega_{n}}{4L_{p}^{n-1}}\right)^{\frac{\Delta}{2}}\cosh\left\{X(r_{a})\right\}}. \label{5.03}
\end{split}
\end{equation}
For the horizon entropy given in Eq.~\eqref{2.11}, we find
\begin{equation}
\begin{split}
    T_{h}\dot{S_{h}} = \dfrac{1}{2\pi r_{a}}\left(n-1\right)\left(1+\dfrac{\Delta}{2}\right)\left(\dfrac{n\Omega_{n}}{4L_{p}^{n-1}}\right)^{1+\frac{\Delta}{2}}r_{a}^{(n-1)\left(1+\frac{\Delta}{2}\right)-1} \dot{r}_{a}\cosh\left\{X(r_{a})\right\}. \label{5.05}
\end{split}
\end{equation}
Substituting Eq.~\eqref{5.03} into Eq.~\eqref{5.05} and simplifying, we obtain
\begin{equation}
    T_{h}\dot S_{h} = n\Omega_{n} r_{a}^{n} H(\rho+P). \label{5.06}
\end{equation}
To study the total entropy change, we also consider the entropy of matter enclosed by the apparent horizon. The Gibbs relation is given by \cite{Izquierdo:2006mt}
\begin{equation}
    T_{m} dS_{m} = Vd\rho+(\rho+P)dV. \label{5.07}
\end{equation}
We assume quasi-static thermal equilibrium between the horizon and the bulk matter at late times, so that $T_{h} = T_{m}$ \cite{Izquierdo:2005ku}. Using the continuity equation, Eq.~\eqref{5.07} becomes
\begin{equation}
    T_{h}\dot{S}_{m} = n\Omega_{n} r_{a}^{n-1}(\rho+P)\dot{r}_{a} - n\Omega_{n} r_{a}^{n} H(\rho+P). \label{5.09}
\end{equation}
Combining Eqs.~\eqref{5.06} and \eqref{5.09}, and substituting for $\dot{r}_{a}$ from Eq.~\eqref{5.03}, we have for the total entropy change,
\begin{equation}
\begin{split}
    &T_{h}(\dot{S}_{h}+\dot{S}_{m}) = \dfrac{8\pi L_{p}^{n-1}}{n-1}\left[\dfrac{n\Omega_{n}r_{a}^{3+(n-1)\left(1-\frac{\Delta}{2}\right)}(\rho +P)^{2}H}{\left(1+\dfrac{\Delta}{2}\right)\left(\dfrac{n\Omega_{n}}{4L_{p}^{n-1}}\right)^{\frac{\Delta}{2}} \cosh\left\{X(r_{a})\right\}}\right]. \label{5.11}
\end{split}
\end{equation}
For an expanding universe, the right-hand side of Eq.~\eqref{5.11} is non-negative throughout the considered parameter range $\Delta\in(-1,1]$. Therefore, the GSL is satisfied within the present framework under the assumed thermal equilibrium conditions.

\section{Density contrast evolution equation} \label{s6}

In this section, we derive the density contrast evolution equation for the present model in a $(3+1)$-dimensional spatially flat universe containing only matter. The density contrast for the matter component is defined as \cite{Ziaie:2019gan}
\begin{equation}
    \delta_{m}=\dfrac{\rho_{m}^{c}}{\rho_{m}} - 1 = \dfrac{\delta \rho_{m}}{\rho_{m}}, \label{6.01}
\end{equation}
where $\rho_{m}^{c} = \rho_{m} + \delta \rho_{m}$ is the perturbed matter energy density. We consider a spherical region containing a small overdensity relative to the homogeneous background universe. The matter density and expansion rate inside this perturbed region are denoted by $\rho_{m}^{c}$ and $H^{c}$, respectively. The corresponding conservation equation is given by $\dot{\rho}_{m}^{c} + 3H^{c}\rho_{m}^{c} = 0$ \cite{Sheykhi:2022gzb}. Taking the first and second time derivatives of Eq.~\eqref{6.01} and using the conservation equation, we obtain
\begin{equation}
    \dot{\delta}_{m} = 3(1 + \delta_{m})(H - H^{c}), \label{6.03}
\end{equation}
\begin{equation}
    \ddot{\delta}_{m} - \dfrac{\dot{\delta}_{m}^{2}}{1 + \delta_{m}} = 3\left(\dot{H} - \dot{H^{c}}\right)\left(1 + \delta_{m}\right). \label{6.04}
\end{equation}
Matter density perturbations have been investigated in cosmological models based on Kaniadakis \cite{Lambiase:2023ryq} and Barrow entropy \cite{Sheykhi:2022gzb}. Here, we extend the analysis to the present model. Since we consider a matter-only universe, the pressure term in Eq.~\eqref{4.15} vanishes, that is, $P=0$. Therefore, from the modified first and second Friedmann equations (\eqref{3.14} and \eqref{4.15}, respectively), we obtain
\begin{equation}
\begin{split}
    \dfrac{\ddot{a}}{a}& = H^{2} - \dfrac{8\pi L_{p}^{2} \rho_{m}}{(2+\Delta)}H^{\Delta}\left(\dfrac{L_{p}^{2}}{\pi}\right)^{\frac{\Delta}{2}}\operatorname{sech}\left\{X(H)\right\}\\
    & = F(\rho_{m}). \label{6.07}
\end{split}
\end{equation}
Following Ref.~\cite{Sheykhi:2022gzb}, we assume that the same form of acceleration equation \eqref{6.07} applies to the spherical perturbed region, characterised by the scale factor $a^{c}$ and the expansion rate $H^{c}$. Thus, we have
\begin{equation}
\begin{split}
    \dfrac{\ddot{a}^{c}}{a^{c}}& = \left(H^{c}\right)^{2} - \dfrac{8\pi L_{p}^{2}\rho_{m}^{c}}{(2+\Delta)}\left(H^{c}\right)^{\Delta}\left(\dfrac{L_{p}^{2}}{\pi}\right)^{\frac{\Delta}{2}}\operatorname{sech}\left\{X(H^{c})\right\}\\
    & = F(\rho_{m}^{c}). \label{6.08}
\end{split}
\end{equation}
Since $\rho_{m}^{c}=\rho_{m}(1+\delta_{m})$, we expand $F(\rho_{m}^{c})$ around $\rho_{m}$ and retain only terms up to first order in $\delta_{m}$. Using Eqs.~\eqref{6.03} and \eqref{6.04}, we obtain in the linear regime \cite{Sheykhi:2022gzb},
\begin{equation}
    \ddot{\delta}_{m} + 2H\dot{\delta}_{m} + 3\rho_{m}\dfrac{dF}{d\rho_{m}}\delta_{m} = 0.
    \label{6.12}
\end{equation}
Changing the independent variable from cosmic time to redshift, we obtain
\begin{equation}
    \frac{d^{2}\delta_{m}}{dz^{2}} + \left[\frac{1}{H}\frac{dH}{dz} - \frac{1}{1+z}\right]\frac{d\delta_{m}}{dz} + \frac{3\rho_{m}}{(1+z)^{2}H^{2}}\frac{dF}{d\rho_{m}}\delta_{m} = 0. \label{6.13}
\end{equation}
Using Eq.~\eqref{6.07} and the modified first Friedmann equation \eqref{3.14}, we find the following derivative of $F(\rho_{m})$:
\begin{equation}
\begin{split}
    \dfrac{dF}{d\rho_{m}} =& \dfrac{-4\pi L_{p}^{2}}{3}\left(\dfrac{2}{2 + \Delta}\right)\left(\dfrac{L_{p}^{2}}{\pi}\right)^{\frac{\Delta}{2}}\dfrac{H^{\Delta}}{\cosh\left\{X(H)\right\}} - \dfrac{\left(8\pi L_{p}^{2}\right)^{2}}{3}\left(\dfrac{\Delta}{(2 + \Delta)^{2}}\right)\left(\dfrac{L_{p}^{2}}{\pi}\right)^{\Delta}\dfrac{\rho_{m}H^{2\Delta - 2}}{\cosh^{2}\left\{X(H)\right\}}\\
    &- \dfrac{\left(8\pi L_{p}^{2}\right)^{2}}{3} \left(\dfrac{K \pi}{L_{p}^{2}H^{2}}\right) \left(\dfrac{1}{2 - \Delta}\right)\left(\dfrac{L_{p}^{2}}{\pi}\right)^{\frac{\Delta}{2}} \dfrac{\rho_{m}H^{\Delta - 2}\tanh\left\{X(H)\right\}}{\cosh^{2}\left\{X(H)\right\}} \label{6.16}
\end{split}
\end{equation}
We introduce the normalised Hubble parameter $h=H/H_0$ and define $\gamma = K(\pi / L_{p}^{2}H_{0}^{2})^{1 + \frac{\Delta}{2}}$, so that $X = \gamma h^{-\left(2 + \Delta\right)}$. Expressing the integration limits in Eq.~\eqref{3.14} in terms of $X$, with the lower limit $X(x=0)=0$ and the upper limit $X(h)$, we obtain
\begin{equation}
    \rho_{m}=\dfrac{3H^{2}}{8\pi L_{p}^{2}}\left(\dfrac{2+\Delta}{2-\Delta}\right)\left[\left(\dfrac{\pi}{L_{p}^{2}}\right)^{\frac{\Delta}{2}}H^{-\Delta}\right]\left[\cosh\left\{X(h)\right\} - H^{\Delta-2}\int_{0}^{X}\dfrac{\sinh
    \left\{X\right\}}{\left[x(X)\right]^{\Delta-2}}d\left\{X\right\}\right]. \label{6.17}
\end{equation}
From the continuity equation for matter, $\dot{\rho}_{m}+3H\rho_{m}=0$, we obtain $\rho_{m}=\rho_{m,0}(1+z)^{3}$, where $\rho_{m,0}$ is the present-day matter energy density. Evaluating Eq.~\eqref{6.17} at $z=0$, where $h=1$, gives
\begin{equation}
    \rho_{m,0} = \dfrac{3H_{0}^{2}}{8\pi L_{p}^{2}}\left(\dfrac{2+\Delta}{2-\Delta}\right)\left[\left(\dfrac{\pi}{L_{p}^{2}}\right)^{\frac{\Delta}{2}}H_{0}^{-\Delta}\right]\left[\cosh\left\{\gamma\right\} - \int_{0}^{\gamma}\left(\dfrac{X}{\gamma}\right)^{\frac{\Delta-2}{\Delta+2}}\sinh
    \left\{X\right\}d\left\{X\right\}\right]. \label{6.18}
\end{equation}
For convenience, we denote the bracketed term in Eq.~\eqref{6.18} by
\begin{equation}
    F(\gamma) = \cosh\left\{\gamma\right\} - \int_{0}^{\gamma}\left(\dfrac{X}{\gamma}\right)^{\frac{\Delta-2}{\Delta+2}}\sinh \left\{X\right\}d\left\{X\right\}.
\end{equation}
Substituting Eq.~\eqref{6.16} into Eq.~\eqref{6.13}, and using Eq.~\eqref{6.18} in the resulting expression, we obtain the density contrast evolution equation for the present model in a $(3+1)$-dimensional spatially flat universe:
\begin{equation}
\begin{split}
    &\dfrac{d^{2}\delta_{m}}{d z^{2}} + \left[\dfrac{1}{h}\dfrac{dh}{dz} - \dfrac{1}{1 + z}\right]\dfrac{d\delta_{m}}{dz} - \left[\dfrac{3h^{\Delta - 2}}{2 - \Delta}\dfrac{F(\gamma)}{\cosh{\left\{X\right\}}}(1 + z)\right]\delta_{m} \\
    &+ \left[\dfrac{9h^{\Delta - 4}}{(2 - \Delta)^{2}}\left(\dfrac{F(\gamma)}{\cosh\left\{X\right\}}\right)^{2}\left(\dfrac{\gamma (2 + \Delta)}{h^{2}}\tanh{\left\{X\right\}} + \Delta h^{\Delta}\right)(1 + z)^{4}\right]\delta_{m} = 0. \label{6.20}
\end{split}
\end{equation}
For $\Delta=0$ and $K\to0$, the Barrow and Kaniadakis corrections vanish, and we obtain
\begin{equation}
\begin{split}
    &\dfrac{d^{2}\delta_{m}}{d z^{2}} + \left[\dfrac{1}{h}\dfrac{dh}{dz} - \dfrac{1}{1 + z}\right]\dfrac{d\delta_{m}}{dz} - \left[\dfrac{3}{2h^{2}}(1 + z)\right]\delta_{m} = 0, \label{6.22}
\end{split}
\end{equation}
which is the same result as that obtained in standard cosmology \cite{Abramo:2007iu}. 

\section{Parameter Inference} \label{s7}

We constrain the free parameters $(\gamma,~\Delta,~H_{0},~M,~r_{d},~\sigma_{8}^{0})$ and investigate the cosmological evolution of a $(3+1)$-dimensional spatially flat, matter-only universe with Barrow-corrected horizon entropy within Kaniadakis statistics. From Eqs.~\eqref{4.15} and \eqref{3.14}, we have
\begin{equation}
    \dot{H} = -\dfrac{8\pi L_{p}^{2} \rho_{m}}{(2+\Delta)}H^{\Delta}\left(\dfrac{L_{p}^{2}}{\pi}\right)^{\frac{\Delta}{2}}\operatorname{sech}\left\{X(H)\right\}. \label{7.02}
\end{equation}
Using Eqs.~\eqref{6.18} and \eqref{7.02}, we obtain the following evolution equation in terms of redshift:
\begin{equation}
    \dfrac{dh}{dz}=\dfrac{3h^{\Delta-1}}{2-\Delta}\operatorname{sech}\left\{\gamma h^{-\left(2+\Delta\right)}\right\}F(\gamma)\left(1+z\right)^{2}. \label{7.03}
\end{equation}

We adopt uniform priors for the free parameters within the ranges $0<\gamma<2$, $-1<\Delta<1$, $40<H_0<100$, $-30<M<-10$, $130<r_d<160$, and $0<\sigma_8^0<2$. For each sampled point in the parameter space, the coupled system of Eqs.~\eqref{7.03} and \eqref{6.20} is solved numerically using the RK45 method.

To determine the initial conditions for the density contrast evolution equation, we first consider the asymptotic form of Eq.~\eqref{6.20}. We then find the asymptotic solution as $\delta(z) = (1 + z)^{s}$, where the exponent $s$ corresponding to the growing mode is given by
\begin{equation}
    s=\frac{1-2\Delta-\sqrt{(2\Delta-1)^{2}+24(1+\Delta)}}{2(2-\Delta)}. \label{7.04}
\end{equation}
The coupled equations are integrated from $z_{i} = 100$ to $z = 0$ \cite{Gomez-Valent:2018nib} to determine $\delta_{m}(0)$ and $\delta'_{m}(0)$.

To constrain the model parameters, we construct a joint chi-square likelihood using 57  OHD measurements \cite{Sharov:2018yvz}, 1701 SNe Ia data points from the Pantheon+ data set \cite{Scolnic:2021amr, Brout:2022vxf}, 13 BAO measurements from the DESI DR2 data release \cite{DESI:2025zpo, DESI:2025zgx}, and 63 $f\sigma_{8}$ measurements from RSD \cite{Kazantzidis:2018rnb}. For the OHD likelihood, the theoretical Hubble parameter is calculated as $H(z)=H_{0}h(z)$. For the Pantheon+ likelihood, we calculate the luminosity distance, which for a spatially flat universe is given by
\begin{equation}
    d_{L}(z)=c(1+z)\int_{0}^{z}\dfrac{dz'}{H(z')},
\end{equation}
where $c$ is the speed of light. The corresponding theoretical apparent magnitude is obtained from
\begin{equation}
    m(z) = 5\log_{10}\left(\dfrac{d_{L}(z)}{\mathrm{Mpc}}\right) + 25 + M,
\end{equation}
where $M$ is the absolute magnitude of Type Ia supernovae. For the BAO likelihood, the transverse comoving distance and the radial comoving distance in a spatially flat universe are given by
\begin{equation}
    D_{M}(z)=c\int_{0}^{z}\dfrac{dz'}{H(z')},
\end{equation}
and
\begin{equation}
    D_{H}(z)=\dfrac{c}{H(z)},
\end{equation}
respectively. For isotropic BAO measurements we use the volume-averaged comoving distance
\begin{equation}
    D_{V}(z)=\left[zD_{M}^{2}(z)D_{H}(z)\right]^{1/3}.
\end{equation}
The theoretical quantities used in the BAO likelihood are $D_{M}/r_{d}$, $D_{H}/r_{d}$, and $D_{V}/r_{d}$, where $r_{d}$ is the sound horizon at the drag epoch. For the RSD likelihood, we consider the growth observable $f\sigma_{8}(z)$, where the growth rate of matter perturbations $f(z)$ is defined in terms of the normalised growth factor $D(z)=\delta_{m}(z)/\delta_{m}(0)$ as
\begin{equation}
    f(z) = -(1+z)\dfrac{D'(z)}{D(z)}.
\end{equation}
The amplitude of matter fluctuations on the scale of $8\,h^{-1}\mathrm{Mpc}$ is expressed as
\begin{equation}
    \sigma_8(z)=\sigma_{8}^{0}D(z),
\end{equation}
where $\sigma_{8}^{0}$ denotes its present-day value. The corresponding growth observable is given by
\begin{equation}
    f\sigma_{8}(z)=-(1+z)\sigma_{8}^{0}D'(z). \label{fsigma8}
\end{equation}

\begin{figure}[H]
    \centering
    \includegraphics[width=0.95\textwidth]{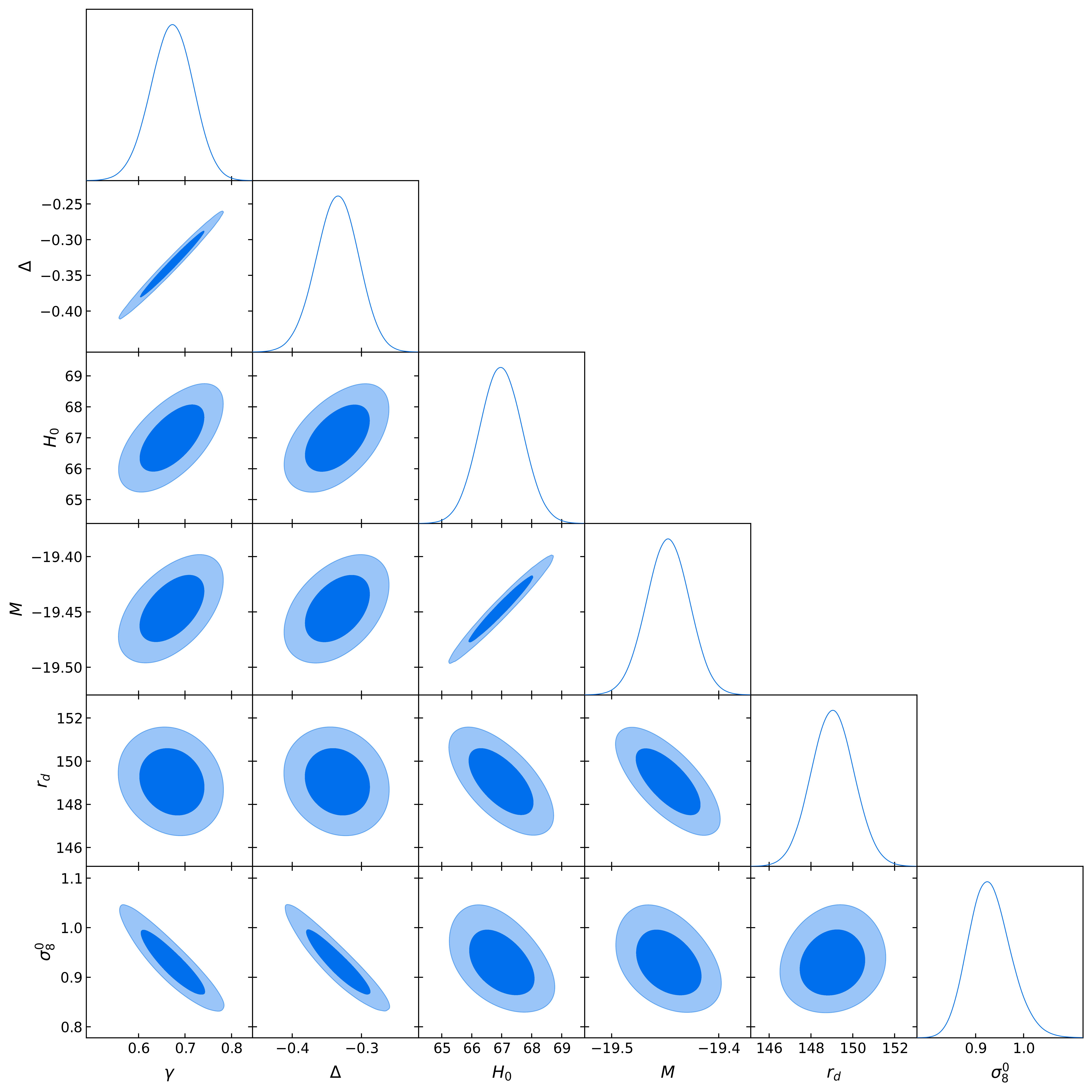}
    \caption{The 1D posterior distribution for the inferred parameters and their 2D contour plots showing 1$\sigma$ and 2$\sigma$ confidence levels constrained against the combined likelihood of the OHD, Pantheon+, DESI DR2, and RSD data}
    \label{GetDist plot}
\end{figure}

The posterior distribution of the model parameters is obtained from Bayes' theorem \cite{Trotta:2008qt, Padilla:2019mgi}. We used an affine-invariant ensemble sampler for MCMC proposed by Goodman and Weare \cite{Goodman:2010dyf}; the implementation of the open-source Python package emcee can be found in \cite{Foreman-Mackey:2012any}. We used 120 walkers and 90000 steps for our MCMC run. The first 30000 steps of the generated chain were discarded as burn-in, and the rest of the samples were used to evaluate the posterior distribution of the model parameters. The results are shown in Figure \ref{GetDist plot} and Table \ref{Best fit values}.
\begin{table}[H]
\centering
\renewcommand{\arraystretch}{1.5}
\setlength{\tabcolsep}{40pt}
\begin{tabular}{ll}
\hline
\hline
Data & OHD + PP + DESI DR2 + RSD \\
\hline
$\gamma$ & $0.672 \pm 0.046$ \\
$\Delta$ & $-0.335 \pm 0.031$ \\
$H_{0}$ & $66.98 \pm 0.71$ \\
$M$ & $-19.447 \pm 0.020$ \\
$r_{d}$ & $149.0 \pm 1.0$ \\
$\sigma_{8}^{0}$ & $0.930^{+0.038}_{-0.048}$ \\ 
\hline
\hline
\end{tabular}
\caption{Median values and 1$\sigma$ confidence intervals for the inferred parameters constrained against the OHD, Pantheon+ (PP), DESI DR2, and RSD data.}
\label{Best fit values}
\end{table}
The parameter $\gamma=K\left(\pi/L_{p}^{2}H_{0} ^{{2}}\right)^{1+\frac{\Delta}{2}}$, estimated as $0.672\pm0.046$ constrains the Kaniadakis parameter $K$ outside its standard Boltzmann-Gibbs limit $(K \to 0)$, implying a preference for $K$-deformed statistics. The Barrow index $\Delta=-0.335\pm0.031$ is constrained to a negative value, $-0.335\pm0.031$, which is different from the Bekenstein-Hawking limit $(\Delta=0)$, implying quantum gravitational deformations on the Hubble horizon. Our result for the Barrow index falls within the broader consideration of the parameter space $(-1, 1]$ mentioned in the introduction. We obtained $H_0=(66.98 \pm 0.71)$ km/s/Mpc, which is closer to the early-universe Planck estimates.

\section{Evolution of Cosmological Parameters} \label{s8}

In this section, we present the evolution of the Hubble parameter, the $f\sigma_{8}$ growth observable, the deceleration parameter, and the age of the universe against redshift. We also examine the statefinder $(r,s)$ pair. 

\begin{figure}[H]
    \centering
    \includegraphics[width=0.7\textwidth]{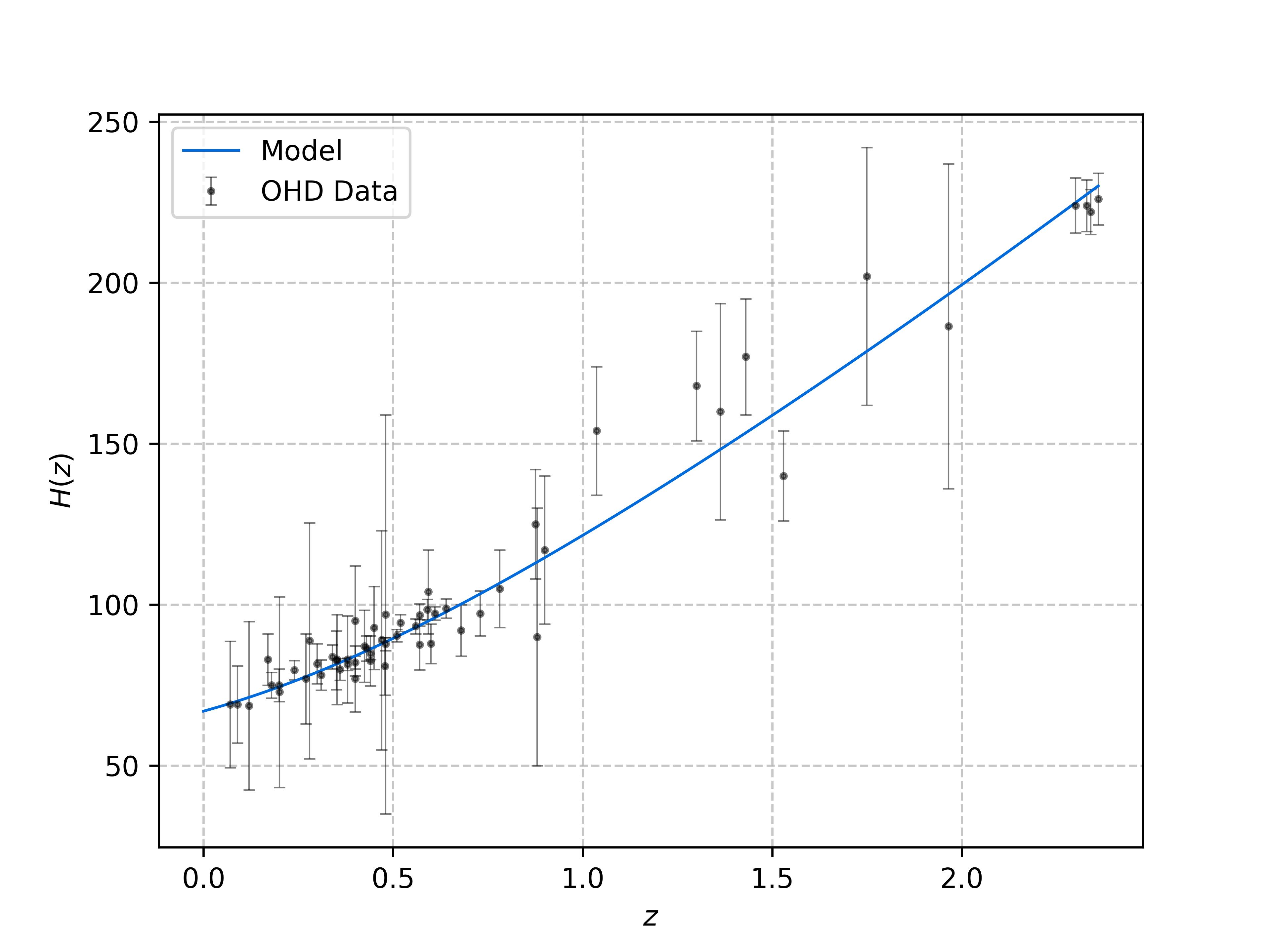}
    \caption{Evolution of the Hubble parameter against redshift using the best-fit model parameters constrained against OHD, Pantheon+, DESI DR2, and RSD data.}
    \label{H vs z}
\end{figure}

\begin{figure}[H]
    \centering
    \includegraphics[width=0.7\textwidth]{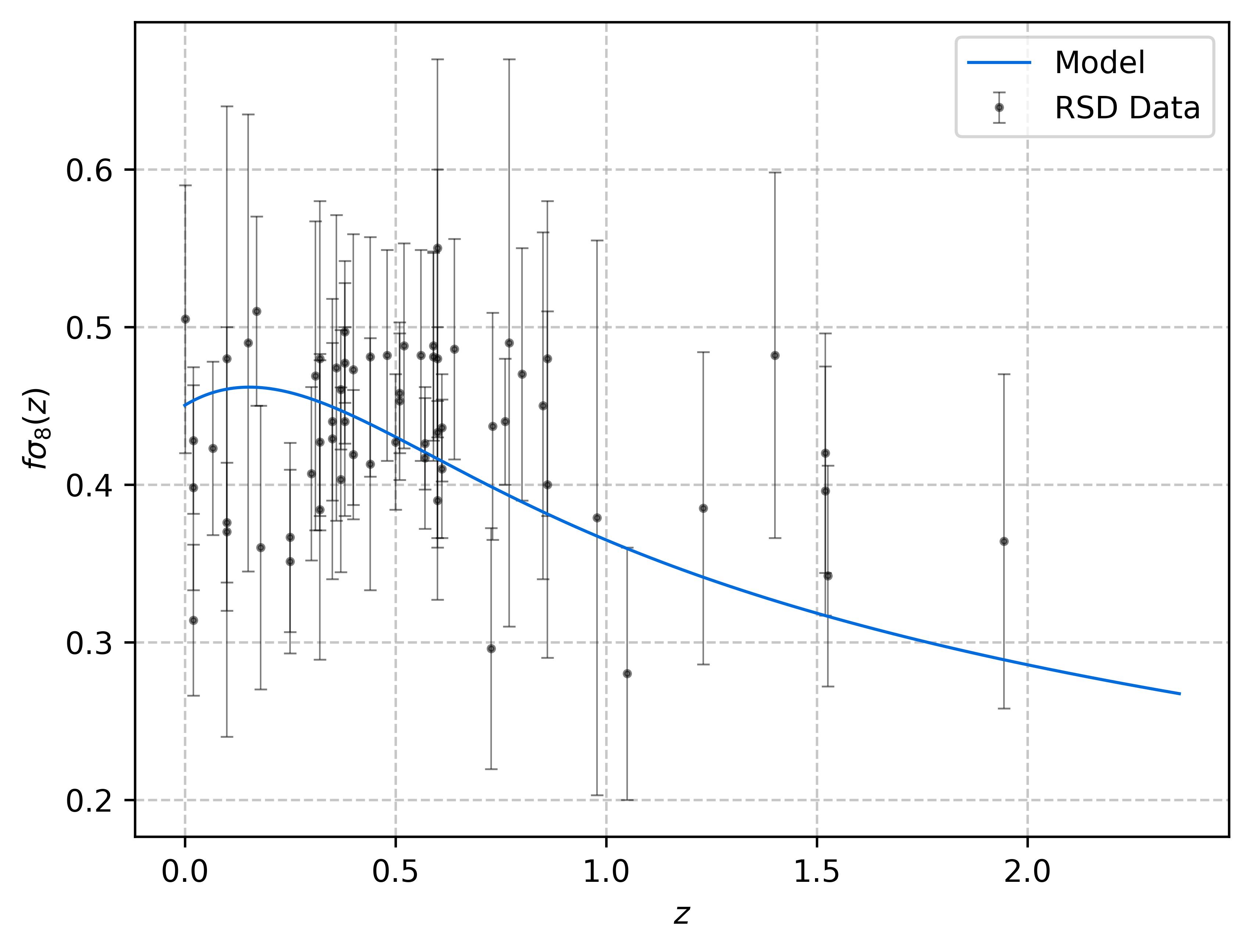}
    \caption{Evolution of the growth rate observable against redshift using the best-fit model parameters constrained against OHD, Pantheon+, DESI DR2, and RSD data.}
    \label{fsigma8 vs z}
\end{figure}

\begin{figure}[H]
    \centering
    \includegraphics[width=0.7\textwidth]{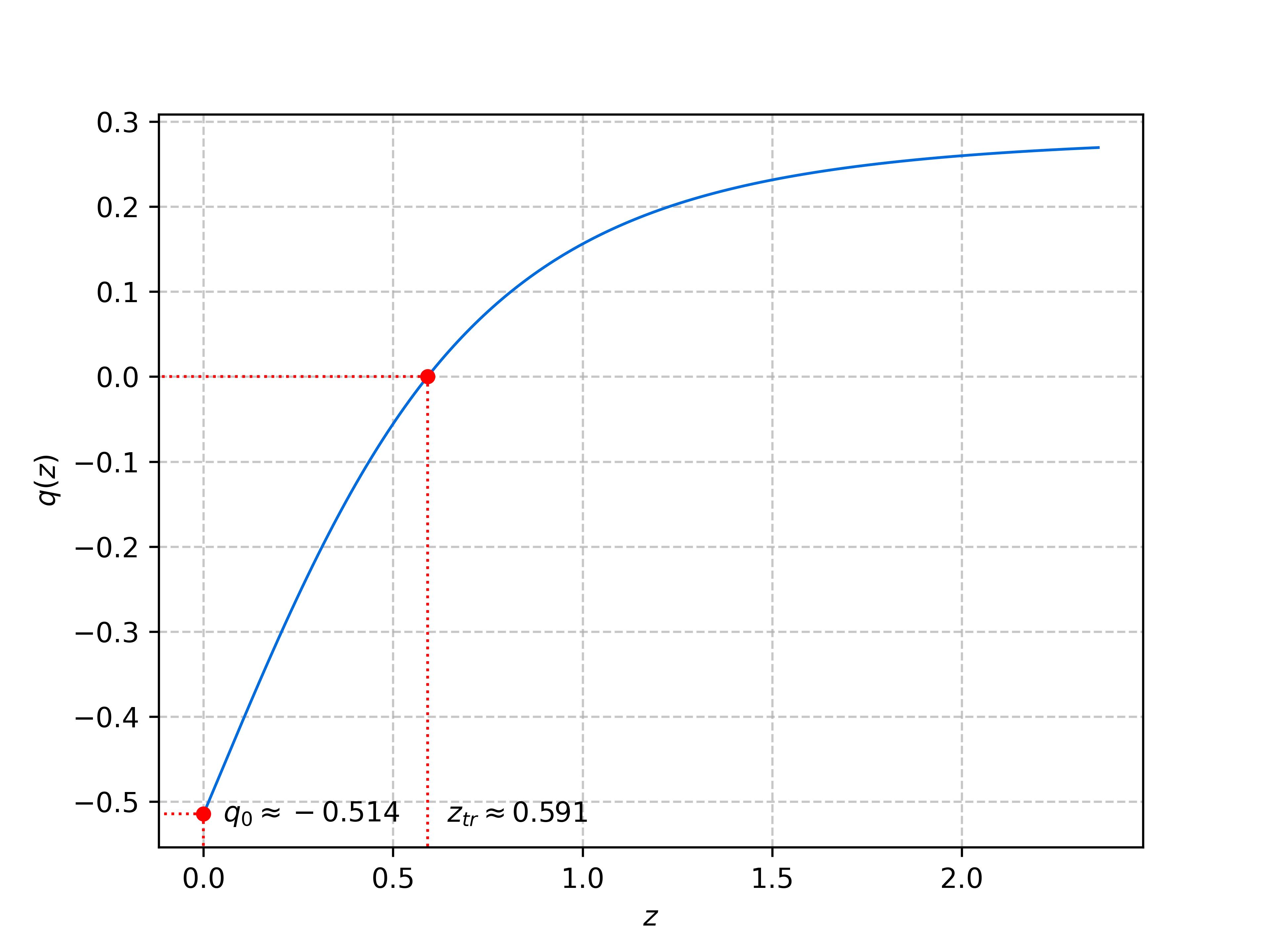}
    \caption{Evolution of the deceleration parameter against redshift using the best-fit model parameters constrained against OHD, Pantheon+, DESI DR2, and RSD data.}
    \label{q vs z}
\end{figure}

\begin{figure}[H]
    \centering
    \includegraphics[width=0.7\textwidth]{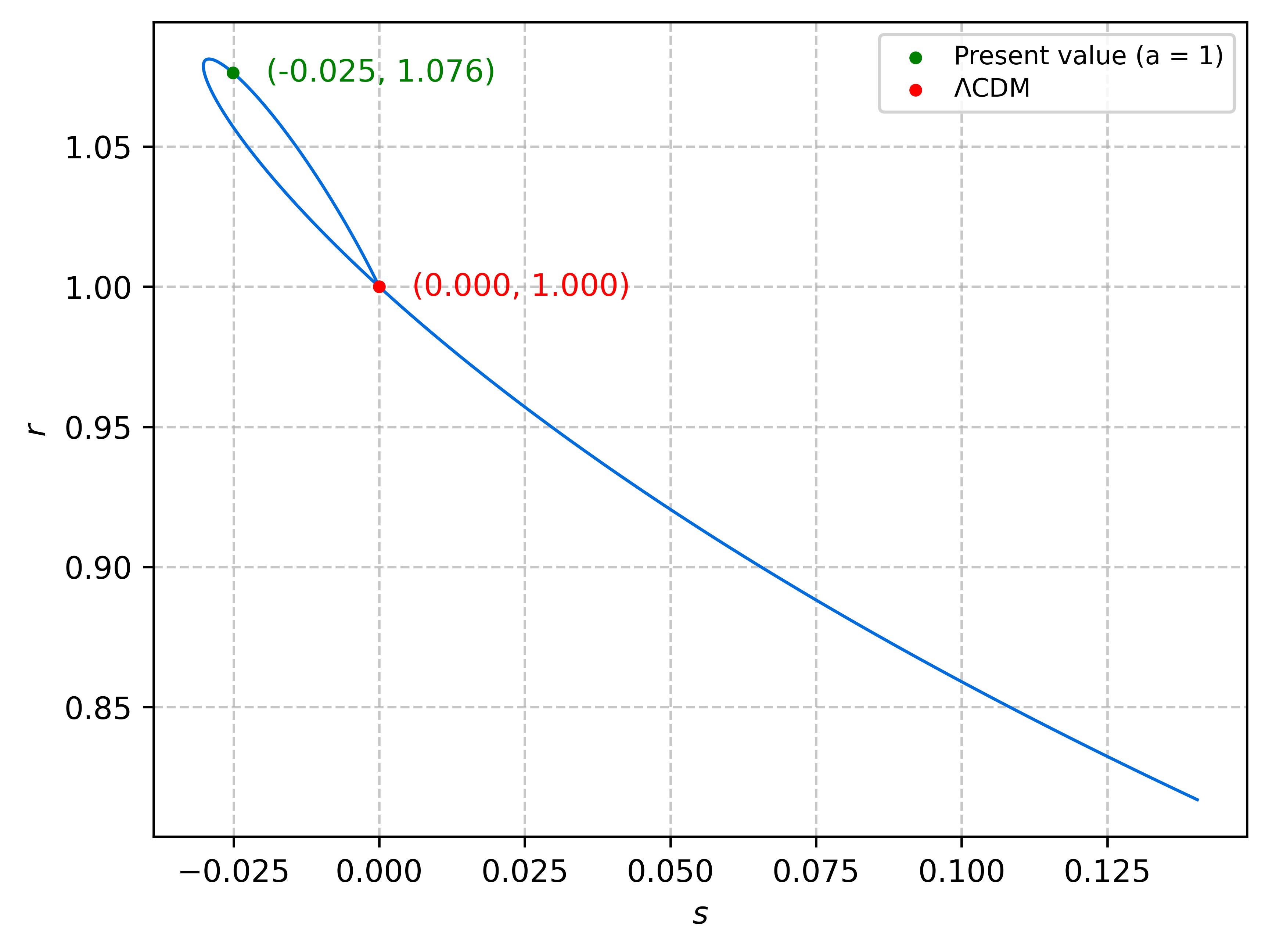}
    \caption{Statefinder $r-s$ plot using the best-fit model parameters constrained against OHD, Pantheon+, DESI DR2, and RSD data.}
    \label{r-s plot}
\end{figure}

Solving Eq.~\eqref{7.03} with the best-fit model parameters, we obtain the evolution of the Hubble parameter. Fig.~\ref{H vs z} shows the expansion history and how it fits the OHD data.
Solving the density contrast equation \eqref{6.20} with the best-fit parameters, we plot $f\sigma_{8}$ against redshift using Eq.~\eqref{fsigma8}. Fig.~\ref{fsigma8 vs z} shows the behaviour of the predicted growth rate observable with increasing redshift and how the model predictions compare with the RSD data.
The age of the universe can be obtained from the definition of the Hubble parameter, $H=\dot{a}/a$, which can be rearranged as $dt/da = (aH(a))^{-1}$ to obtain
\begin{equation}
    t_{0} - t_{B} = \int_{0}^{1}\dfrac{1}{aH}da, 
\end{equation}
where $t_{0}$ is the present age of the universe and $t_{B}$ is the time corresponding to the Big Bang, which is set to zero. The present age of the universe predicted by the model is $14.354$ Gyr, which is higher than the Planck 2018 estimate of $(13.797 \pm 0.023)$ Gyr within the $\Lambda$CDM framework \cite{Planck:2018vyg}.
\begin{figure}[H]
    \centering
    \includegraphics[width=0.7\textwidth]{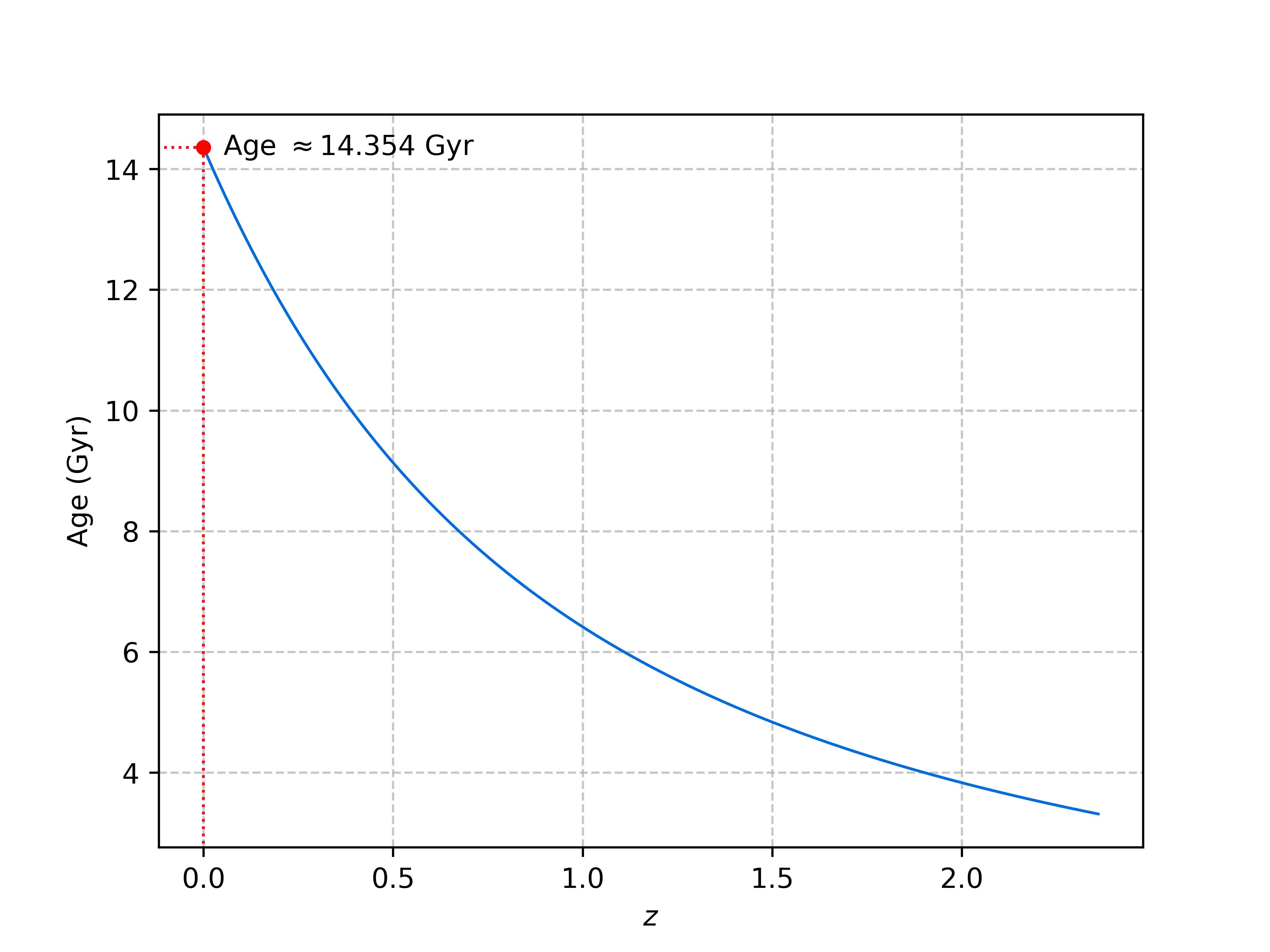}
    \caption{Age of the universe against redshift using the best-fit model parameters constrained against OHD, Pantheon+, DESI DR2, and RSD data.}
    \label{Age vs z}
\end{figure}
The deceleration parameter $(q)$ is a dimensionless parameter that characterises the accelerating and decelerating phases of the universe. It takes a negative value for accelerated expansion and a positive value for deceleration. It is expressed as
\begin{equation}
    q = -1 - \dfrac{\dot{H}}{H^{2}},
\end{equation}
which can be expressed in terms of redshift as
\begin{equation}
    q = -1 + \dfrac{(1+z)}{h}\dfrac{dh}{dz}.
\end{equation}
The evolution of the deceleration parameter with redshift, illustrated in Fig.~\ref{q vs z}, shows a smooth transition from decelerated to accelerated expansion. The transition point is characterised by the redshift $z_{tr}=0.591$, which is comparable to the range $0.5\text{--}0.73$ obtained for the $\Lambda$CDM model \cite{Alam:2004jy, Colistete:2007xi}. The present value of the deceleration parameter is found to be $q_{0} = -0.514$, which is close to the value predicted by the $\Lambda$CDM model \cite{Planck:2018vyg}.
The statefinder diagnostic pair is a geometric tool introduced by Sahni et al. \cite{Sahni:2002fz}. The parameter $r(z)$, involving the third derivative of the scale factor, $\dddot{a}$, extends the hierarchy of geometrical cosmological parameters beyond the Hubble parameter $H$ and the deceleration parameter $q$, while $s(z)$ is defined in terms of $q$ and $r$:
\begin{align}
    &r = \dfrac{\dddot{a}}{aH^{3}},\\[10pt]
    &s = \dfrac{r-1}{3\left(q-\dfrac{1}{2}\right)}.
\end{align}
These variables can be expressed in terms of the dimensionless Hubble parameter $h=H/H_{0}$ by introducing the e-folding number $x=\ln(a)$ \cite{Mathew:2012md}:
\begin{align}
    &r = \dfrac{1}{2h^{2}}\dfrac{d^{2}h^{2}}{dx^{2}} + \dfrac{3}{2h^{2}}\dfrac{dh^{2}}{dx} + 1,\\[10pt]
    &s = -\dfrac{\dfrac{1}{2h^{2}}\dfrac{d^{2}h^{2}}{dx^{2}} + \dfrac{3}{2h^{2}}\dfrac{dh^{2}}{dx}}{\dfrac{3}{2h^{2}}\dfrac{dh^{2}}{dx} + \dfrac{9}{2}}.
\end{align}
As illustrated in Fig.~\ref{r-s plot}, the $r-s$ curve passes through the fixed $\Lambda$CDM point $(r, s)=(1,0)$ in the past, exhibits a turnover, and eventually converges back to the $\Lambda$CDM point after the present time \cite{Carrasco:2023imi, Sardar:2024bfv}. The present value $(r_{0},s_{0})$ of the model is obtained as $(1.076,-0.025)$.

\section{Discussion and Conclusions} \label{s9}

The objective of this research was to consider the Barrow correction to the horizon entropy within Kaniadakis statistics and to study the late-time accelerated expansion of a universe consisting only of matter. We derived the modified Friedmann equations for an $(n+1)$-dimensional non-flat universe using both the law of emergence and the first law of thermodynamics. We also verified that the model satisfies the generalised second law of thermodynamics. The density contrast evolution equation was derived for a $(3+1)$-dimensional flat universe to study linear matter density perturbations within the model.

To validate the theoretical formulation and study the model's implications, an MCMC analysis was performed using the emcee library within a Bayesian framework. The model parameters were constrained using the joint likelihood of the OHD, Pantheon+, DESI DR2, and RSD datasets. The parameter $\gamma$ was constrained to $\gamma=0.672\pm0.046$, indicating a preference for a non-zero Kaniadakis parameter $K$. The Barrow index $\Delta= -0.335 \pm 0.031$ is constrained to a negative value. As mentioned in the introduction, this may indicate a horizon surface with porosity or internal voids instead of a sphereflake construction. Our constraint on the Barrow index falls within the broader consideration of the parameter space $(-1, 1]$ considered for a more general framework \cite{Jizba:2024klq, Luciano:2025hjn}. The present value of the Hubble parameter was constrained as $H_{0}=(66.98 \pm 0.71)$ km/s/Mpc, which leans towards the early-universe Planck estimates; however, late-time data alone are insufficient to address the Hubble tension problem. We need to incorporate Cosmic Microwave Background (CMB) data \cite{Planck:2018vyg} into the parameter inference pipeline to enable a joint constraint across a wide range of redshifts, spanning from the recombination era to the late-time universe. The present-day amplitude of matter density fluctuations is constrained as $\sigma_{8}^{0}=0.930^{+0.038}_{-0.048}$.

The expansion history of the universe predicted by the model was evaluated by examining the Hubble parameter, the deceleration parameter, and the age of the universe against redshift. The growth of linear matter density perturbations was also analysed through the behaviour of the growth observable $f\sigma_{8}$ against redshift. The deceleration parameter shows a smooth transition from decelerated to accelerated expansion at a transition redshift $z_{tr}=0.591$, which falls within the range $0.5\text{--}0.73$ obtained for the $\Lambda$CDM model \cite{Alam:2004jy, Colistete:2007xi}. The present value of the deceleration parameter is found to be $q_{0}=-0.514$, which is comparable to the $\Lambda$CDM value \cite{Planck:2018vyg}, without invoking dark energy. Based on this expansion history, the present age of the universe is obtained as $14.354$ Gyr, which is higher than the Planck 2018 estimate of $(13.797\pm0.023)$ Gyr \cite{Planck:2018vyg}. The statefinder $(r,s)$ diagnostic was also studied, and the present-day (r,s) coordinates were obtained as $(1.076,-0.025)$, which is different from the fixed $\Lambda$CDM point $(1, 0)$.

In summary, the results demonstrate that the Barrow-corrected horizon entropy within Kaniadakis statistics in a matter-only framework can reproduce late-time acceleration and viable expansion history consistent with the considered observational data. Although the present analysis does not address Hubble tension nor $S_{8}$ tension, the model will be further studied using early-universe data.

\bibliographystyle{unsrturl}
\bibliography{Bibilography}

@article{Bekenstein:1973ur,
    author = "Bekenstein, Jacob D.",
    title = "{Black holes and entropy}",
    doi = "10.1103/PhysRevD.7.2333",
    journal = "Phys. Rev. D",
    volume = "7",
    pages = "2333--2346",
    year = "1973"
}

@article{Hawking:1975vcx,
    author = "Hawking, S. W.",
    editor = "Gibbons, G. W. and Hawking, S. W.",
    title = "{Particle Creation by Black Holes}",
    doi = "10.1007/BF02345020",
    journal = "Commun. Math. Phys.",
    volume = "43",
    pages = "199--220",
    year = "1975",
    note = "[Erratum: Commun.Math.Phys. 46, 206 (1976)]"
}

@article{Jacobson:1995ab,
    author = "Jacobson, Ted",
    title = "{Thermodynamics of space-time: The Einstein equation of state}",
    eprint = "gr-qc/9504004",
    archivePrefix = "arXiv",
    reportNumber = "UMDGR-95-114",
    doi = "10.1103/PhysRevLett.75.1260",
    journal = "Phys. Rev. Lett.",
    volume = "75",
    pages = "1260--1263",
    year = "1995"
}

@article{Padmanabhan:2009kr,
    author = "Padmanabhan, T.",
    title = "{Equipartition of energy in the horizon degrees of freedom and the emergence of gravity}",
    eprint = "0912.3165",
    archivePrefix = "arXiv",
    primaryClass = "gr-qc",
    doi = "10.1142/S021773231003313X",
    journal = "Mod. Phys. Lett. A",
    volume = "25",
    pages = "1129--1136",
    year = "2010"
}

@article{Sheykhi:2021fwh,
    author = "Sheykhi, Ahmad",
    title = "{Barrow Entropy Corrections to Friedmann Equations}",
    eprint = "2102.06550",
    archivePrefix = "arXiv",
    primaryClass = "gr-qc",
    doi = "10.1103/PhysRevD.103.123503",
    journal = "Phys. Rev. D",
    volume = "103",
    number = "12",
    pages = "123503",
    year = "2021"
}

@article{KrishnaPB:2020sne,
    author = "B, Krishna P. and Mathew, Titus K.",
    title = "{Emergence of cosmic space and the maximization of horizon entropy}",
    eprint = "2002.02121",
    archivePrefix = "arXiv",
    primaryClass = "gr-qc",
    doi = "10.1016/j.dark.2024.101451",
    journal = "Phys. Dark Univ.",
    volume = "44",
    pages = "101451",
    year = "2024"
}

@article{Padmanabhan:2012ik,
    author = "Padmanabhan, T.",
    title = "{Emergence and Expansion of Cosmic Space as due to the Quest for Holographic Equipartition}",
    eprint = "1206.4916",
    archivePrefix = "arXiv",
    primaryClass = "hep-th",
    month = "6",
    year = "2012"
}

@article{Cai:2012ip,
    author = "Cai, Rong-Gen",
    title = "{Emergence of Space and Spacetime Dynamics of Friedmann-Robertson-Walker Universe}",
    eprint = "1207.0622",
    archivePrefix = "arXiv",
    primaryClass = "gr-qc",
    doi = "10.1007/JHEP11(2012)016",
    journal = "JHEP",
    volume = "11",
    pages = "016",
    year = "2012"
}

@article{Sheykhi:2013oac,
    author = "Sheykhi, Ahmad",
    title = "{Friedmann equations from emergence of cosmic space}",
    eprint = "1304.3054",
    archivePrefix = "arXiv",
    primaryClass = "gr-qc",
    doi = "10.1103/PhysRevD.87.061501",
    journal = "Phys. Rev. D",
    volume = "87",
    number = "6",
    pages = "061501",
    year = "2013"
}

@article{Cai:2008gw,
    author = "Cai, Rong-Gen and Cao, Li-Ming and Hu, Ya-Peng",
    title = "{Hawking Radiation of Apparent Horizon in a FRW Universe}",
    eprint = "0809.1554",
    archivePrefix = "arXiv",
    primaryClass = "hep-th",
    doi = "10.1088/0264-9381/26/15/155018",
    journal = "Class. Quant. Grav.",
    volume = "26",
    pages = "155018",
    year = "2009"
}

@article{Cai:2005ra,
    author = "Cai, Rong-Gen and Kim, Sang Pyo",
    title = "{First law of thermodynamics and Friedmann equations of Friedmann-Robertson-Walker universe}",
    eprint = "hep-th/0501055",
    archivePrefix = "arXiv",
    doi = "10.1088/1126-6708/2005/02/050",
    journal = "JHEP",
    volume = "02",
    pages = "050",
    year = "2005"
}

@article{Tsallis:1987eu,
    author = "Tsallis, Constantino",
    title = "{Possible Generalization of Boltzmann-Gibbs Statistics}",
    reportNumber = "CBPF-NF-062-87",
    doi = "10.1007/BF01016429",
    journal = "J. Statist. Phys.",
    volume = "52",
    pages = "479--487",
    year = "1988"
}

@article{Tsallis:2012js,
    author = "Tsallis, Constantino and Cirto, Leonardo J. L.",
    title = "{Black hole thermodynamical entropy}",
    eprint = "1202.2154",
    archivePrefix = "arXiv",
    primaryClass = "cond-mat.stat-mech",
    doi = "10.1140/epjc/s10052-013-2487-6",
    journal = "Eur. Phys. J. C",
    volume = "73",
    pages = "2487",
    year = "2013"
}

@article{Kaniadakis:2001tcv,
    author = "Kaniadakis, G.",
    title = "{Non-linear kinetics underlying generalized statistics}",
    eprint = "cond-mat/0103467",
    archivePrefix = "arXiv",
    doi = "10.1016/s0378-4371(01)00184-4",
    journal = "Physica A",
    volume = "296",
    number = "3-4",
    pages = "405--425",
    year = "2001"
}

@article{Kaniadakis:2002zz,
    author = "Kaniadakis, G.",
    title = "{Statistical mechanics in the context of special relativity}",
    eprint = "cond-mat/0210467",
    archivePrefix = "arXiv",
    doi = "10.1103/PhysRevE.66.056125",
    journal = "Phys. Rev. E",
    volume = "66",
    pages = "056125",
    year = "2002"
}

@article{Barrow:2020tzx,
    author = "Barrow, John D.",
    title = "{The Area of a Rough Black Hole}",
    eprint = "2004.09444",
    archivePrefix = "arXiv",
    primaryClass = "gr-qc",
    doi = "10.1016/j.physletb.2020.135643",
    journal = "Phys. Lett. B",
    volume = "808",
    pages = "135643",
    year = "2020"
}

@article{Moradpour:2020dfm,
    author = "Moradpour, H. and Ziaie, A. H. and Kord Zangeneh, M.",
    title = "{Generalized entropies and corresponding holographic dark energy models}",
    eprint = "2005.06271",
    archivePrefix = "arXiv",
    primaryClass = "gr-qc",
    reportNumber = "80,732",
    doi = "10.1140/epjc/s10052-020-8307-x",
    journal = "Eur. Phys. J. C",
    volume = "80",
    number = "8",
    pages = "732",
    year = "2020"
}

@article{Lymperis:2021qty,
    author = "Lymperis, Andreas and Basilakos, Spyros and Saridakis, Emmanuel N.",
    title = "{Modified cosmology through Kaniadakis horizon entropy}",
    eprint = "2108.12366",
    archivePrefix = "arXiv",
    primaryClass = "gr-qc",
    doi = "10.1140/epjc/s10052-021-09852-9",
    journal = "Eur. Phys. J. C",
    volume = "81",
    number = "11",
    pages = "1037",
    year = "2021"
}

@article{Saridakis:2020cqq,
    author = "Saridakis, Emmanuel N. and Basilakos, Spyros",
    title = "{The generalized second law of thermodynamics with Barrow entropy}",
    eprint = "2005.08258",
    archivePrefix = "arXiv",
    primaryClass = "gr-qc",
    doi = "10.1140/epjc/s10052-021-09431-y",
    journal = "Eur. Phys. J. C",
    volume = "81",
    number = "7",
    pages = "644",
    year = "2021"
}

@article{Saridakis:2020lrg,
    author = "Saridakis, Emmanuel N.",
    title = "{Modified cosmology through spacetime thermodynamics and Barrow horizon entropy}",
    eprint = "2006.01105",
    archivePrefix = "arXiv",
    primaryClass = "gr-qc",
    doi = "10.1088/1475-7516/2020/07/031",
    journal = "JCAP",
    volume = "07",
    pages = "031",
    year = "2020"
}

@article{Luciano:2023roh,
    author = "Luciano, Giuseppe Gaetano",
    title = "{Constraining barrow entropy-based cosmology with power-law inflation}",
    eprint = "2301.12509",
    archivePrefix = "arXiv",
    primaryClass = "gr-qc",
    doi = "10.1140/epjc/s10052-023-11499-7",
    journal = "Eur. Phys. J. C",
    volume = "83",
    number = "4",
    pages = "329",
    year = "2023"
}

@article{Prasanthan:2024xsl,
    author = "Prasanthan, Pranav and Nelleri, Sarath and Poonthottathil, Navaneeth and K, Sreejith E.",
    title = "{Emergence of cosmic space and horizon thermodynamics from Kaniadakis entropy}",
    eprint = "2405.03592",
    archivePrefix = "arXiv",
    primaryClass = "gr-qc",
    doi = "10.1088/1361-6382/ad6e4e",
    journal = "Class. Quant. Grav.",
    volume = "41",
    number = "18",
    pages = "185008",
    year = "2024"
}

@article{Tian:2014ila,
    author = "Tian, David Wenjie and Booth, Ivan",
    title = "{Friedmann equations from nonequilibrium thermodynamics of the Universe: A unified formulation for modified gravity}",
    eprint = "1409.4278",
    archivePrefix = "arXiv",
    primaryClass = "gr-qc",
    doi = "10.1103/PhysRevD.90.104042",
    journal = "Phys. Rev. D",
    volume = "90",
    number = "10",
    pages = "104042",
    year = "2014"
}

@article{Bekenstein:1974ax,
    author = "Bekenstein, Jacob D.",
    title = "{Generalized second law of thermodynamics in black-hole physics}",
    journal = "Phys. Rev. D",
    volume = "9",
    pages = "3292--3300",
    year = "1974",
    doi = "10.1103/PhysRevD.9.3292"
}

@inproceedings{Izquierdo:2006mt,
    author = "Izquierdo, German and Pavon, Diego",
    title = "{The Generalized second law in dark energy dominated universes}",
    booktitle = "{11th Marcel Grossmann Meeting on General Relativity}",
    eprint = "gr-qc/0612092",
    archivePrefix = "arXiv",
    doi = "10.1142/9789812834300_0249",
    pages = "1764--1666",
    month = "12",
    year = "2006"
}

@article{Ziaie:2019gan,
    author = "Ziaie, A. H. and Shabani, H. and Ghaffari, S.",
    title = "{Effects of Rastall parameter on perturbation of dark sectors of the Universe}",
    eprint = "1909.12085",
    archivePrefix = "arXiv",
    primaryClass = "physics.gen-ph",
    doi = "10.1142/S0217732321500826",
    journal = "Mod. Phys. Lett. A",
    volume = "36",
    number = "30",
    pages = "2150082",
    year = "2021"
}

@article{Sheykhi:2022gzb,
    author = "Sheykhi, Ahmad and Farsi, Bita",
    title = "{Growth of perturbations in Tsallis and Barrow cosmology}",
    eprint = "2205.04138",
    archivePrefix = "arXiv",
    primaryClass = "gr-qc",
    doi = "10.1140/epjc/s10052-022-11044-y",
    journal = "Eur. Phys. J. C",
    volume = "82",
    number = "12",
    pages = "1111",
    year = "2022"
}

@article{Trotta:2008qt,
    author = "Trotta, Roberto",
    title = "{Bayes in the sky: Bayesian inference and model selection in cosmology}",
    eprint = "0803.4089",
    archivePrefix = "arXiv",
    primaryClass = "astro-ph",
    doi = "10.1080/00107510802066753",
    journal = "Contemp. Phys.",
    volume = "49",
    pages = "71--104",
    year = "2008"
}

@article{Sharov:2018yvz,
    author = "Sharov, G. S. and Vasiliev, V. O.",
    title = "{How predictions of cosmological models depend on Hubble parameter data sets}",
    eprint = "1807.07323",
    archivePrefix = "arXiv",
    primaryClass = "gr-qc",
    doi = "10.26456/mmg/2018-611",
    journal = "Math. Model. Geom.",
    volume = "6",
    pages = "1--20",
    year = "2018"
}

@article{Scolnic:2021amr,
    author = "Scolnic, Dan and others",
    title = "{The Pantheon+ Analysis: The Full Data Set and Light-curve Release}",
    eprint = "2112.03863",
    archivePrefix = "arXiv",
    primaryClass = "astro-ph.CO",
    doi = "10.3847/1538-4357/ac8b7a",
    journal = "Astrophys. J.",
    volume = "938",
    number = "2",
    pages = "113",
    year = "2022"
}

@article{Brout:2022vxf,
    author = "Brout, Dillon and others",
    title = "{The Pantheon+ Analysis: Cosmological Constraints}",
    eprint = "2202.04077",
    archivePrefix = "arXiv",
    primaryClass = "astro-ph.CO",
    doi = "10.3847/1538-4357/ac8e04",
    journal = "Astrophys. J.",
    volume = "938",
    number = "2",
    pages = "110",
    year = "2022"
}

@article{DESI:2025zpo,
    author = "Abdul Karim, M. and others",
    collaboration = "DESI",
    title = "{DESI DR2 results. I. Baryon acoustic oscillations from the Lyman alpha forest}",
    eprint = "2503.14739",
    archivePrefix = "arXiv",
    primaryClass = "astro-ph.CO",
    reportNumber = "FERMILAB-PUB-25-0167-PPD",
    doi = "10.1103/2wwn-xjm5",
    journal = "Phys. Rev. D",
    volume = "112",
    number = "8",
    pages = "083514",
    year = "2025"
}

@article{DESI:2025zgx,
    author = "Abdul Karim, M. and others",
    collaboration = "DESI",
    title = "{DESI DR2 results. II. Measurements of baryon acoustic oscillations and cosmological constraints}",
    eprint = "2503.14738",
    archivePrefix = "arXiv",
    primaryClass = "astro-ph.CO",
    reportNumber = "FERMILAB-PUB-25-0169-PPD",
    doi = "10.1103/tr6y-kpc6",
    journal = "Phys. Rev. D",
    volume = "112",
    number = "8",
    pages = "083515",
    year = "2025"
}

@article{Kazantzidis:2018rnb,
    author = "Kazantzidis, Lavrentios and Perivolaropoulos, Leandros",
    title = "{Evolution of the $f\sigma_8$ tension with the Planck15/$\Lambda$CDM determination and implications for modified gravity theories}",
    eprint = "1803.01337",
    archivePrefix = "arXiv",
    primaryClass = "astro-ph.CO",
    doi = "10.1103/PhysRevD.97.103503",
    journal = "Phys. Rev. D",
    volume = "97",
    number = "10",
    pages = "103503",
    year = "2018"
}

@article{Gomez-Valent:2018nib,
    author = "G{\'o}mez-Valent, Adri{\`a} and Sol{\`a} Peracaula, Joan",
    title = "{Density perturbations for running vacuum: a successful approach to structure formation and to the $\sigma_8$-tension}",
    eprint = "1801.08501",
    archivePrefix = "arXiv",
    primaryClass = "astro-ph.CO",
    doi = "10.1093/mnras/sty1028",
    journal = "Mon. Not. Roy. Astron. Soc.",
    volume = "478",
    number = "1",
    pages = "126--145",
    year = "2018"
}

@article{Planck:2018vyg,
    author = "Aghanim, N. and others",
    collaboration = "Planck",
    title = "{Planck 2018 results. VI. Cosmological parameters}",
    eprint = "1807.06209",
    archivePrefix = "arXiv",
    primaryClass = "astro-ph.CO",
    doi = "10.1051/0004-6361/201833910",
    journal = "Astron. Astrophys.",
    volume = "641",
    pages = "A6",
    year = "2020",
    note = "[Erratum: Astron.Astrophys. 652, C4 (2021)]"
}

@article{Goodman:2010dyf,
    author = "Goodman, Jonathan and Weare, Jonathan",
    title = "{Ensemble samplers with affine invariance}",
    doi = "10.2140/camcos.2010.5.65",
    journal = "Commun. Appl. Math. Comput. Sc.",
    volume = "5",
    number = "1",
    pages = "65--80",
    year = "2010"
}

@article{Foreman-Mackey:2012any,
    author = "Foreman-Mackey, Daniel and Hogg, David W. and Lang, Dustin and Goodman, Jonathan",
    title = "{emcee: The MCMC Hammer}",
    eprint = "1202.3665",
    archivePrefix = "arXiv",
    primaryClass = "astro-ph.IM",
    doi = "10.1086/670067",
    journal = "Publ. Astron. Soc. Pac.",
    volume = "125",
    pages = "306--312",
    year = "2013"
}

@article{Alam:2004jy,
    author = "Alam, Ujjaini and Sahni, Varun and Starobinsky, A. A.",
    title = "{The Case for dynamical dark energy revisited}",
    eprint = "astro-ph/0403687",
    archivePrefix = "arXiv",
    doi = "10.1088/1475-7516/2004/06/008",
    journal = "JCAP",
    volume = "06",
    pages = "008",
    year = "2004"
}

@article{Colistete:2007xi,
    author = "Colistete, R. and Fabris, J. C. and Tossa, J. and Zimdahl, W.",
    title = "{Bulk Viscous Cosmology}",
    eprint = "0706.4086",
    archivePrefix = "arXiv",
    primaryClass = "astro-ph",
    doi = "10.1103/PhysRevD.76.103516",
    journal = "Phys. Rev. D",
    volume = "76",
    pages = "103516",
    year = "2007"
}

@article{Sahni:2002fz,
    author = "Sahni, Varun and Saini, Tarun Deep and Starobinsky, Alexei A. and Alam, Ujjaini",
    title = "{Statefinder: A New geometrical diagnostic of dark energy}",
    eprint = "astro-ph/0201498",
    archivePrefix = "arXiv",
    doi = "10.1134/1.1574831",
    journal = "JETP Lett.",
    volume = "77",
    pages = "201--206",
    year = "2003"
}

@article{Mathew:2012md,
    author = "Mathew, Titus K. and Suresh, Jishnu and Divakaran, Divya",
    title = "{Modified holographic Ricci dark energy model and statefinder diagnosis in flat universe}",
    eprint = "1207.5886",
    archivePrefix = "arXiv",
    primaryClass = "astro-ph.CO",
    doi = "10.1142/S0218271813500569",
    journal = "Int. J. Mod. Phys. D",
    volume = "22",
    pages = "1350056",
    year = "2013"
}

@article{Luciano:2025hjn,
    author = "Luciano, Giuseppe Gaetano and Paliathanasis, Andronikos and Saridakis, Emmanuel N.",
    title = "{Barrow and Tsallis entropies after the DESI DR2 BAO data}",
    eprint = "2504.12205",
    archivePrefix = "arXiv",
    primaryClass = "gr-qc",
    doi = "10.1088/1475-7516/2025/09/013",
    journal = "JCAP",
    volume = "09",
    pages = "013",
    year = "2025"
}

@article{Abramo:2007iu,
    author = "Abramo, L. R. and Batista, R. C. and Liberato, L. and Rosenfeld, R.",
    title = "{Structure formation in the presence of dark energy perturbations}",
    eprint = "0707.2882",
    archivePrefix = "arXiv",
    primaryClass = "astro-ph",
    doi = "10.1088/1475-7516/2007/11/012",
    journal = "JCAP",
    volume = "11",
    pages = "012",
    year = "2007"
}

@article{Hayward:1993wb,
    author = "Hayward, S. A.",
    title = "{General laws of black hole dynamics}",
    doi = "10.1103/PhysRevD.49.6467",
    journal = "Phys. Rev. D",
    volume = "49",
    pages = "6467--6474",
    year = "1994"
}

@article{Hayward:1998ee,
    author = "Hayward, Sean A. and Mukohyama, Shinji and Ashworth, M. C.",
    title = "{Dynamic black hole entropy}",
    eprint = "gr-qc/9810006",
    archivePrefix = "arXiv",
    doi = "10.1016/S0375-9601(99)00225-X",
    journal = "Phys. Lett. A",
    volume = "256",
    pages = "347--350",
    year = "1999"
}

@article{Izquierdo:2005ku,
    author = "Izquierdo, German and Pavon, Diego",
    title = "{Dark energy and the generalized second law}",
    eprint = "astro-ph/0505601",
    archivePrefix = "arXiv",
    doi = "10.1016/j.physletb.2005.12.040",
    journal = "Phys. Lett. B",
    volume = "633",
    pages = "420--426",
    year = "2006"
}

@article{Komatsu:2016vof,
    author = "Komatsu, Nobuyoshi",
    title = "{Cosmological model from the holographic equipartition law with a modified R{\'e}nyi entropy}",
    eprint = "1611.04084",
    archivePrefix = "arXiv",
    primaryClass = "gr-qc",
    doi = "10.1140/epjc/s10052-017-4800-2",
    journal = "Eur. Phys. J. C",
    volume = "77",
    number = "4",
    pages = "229",
    year = "2017"
}

@article{Sheykhi:2018dpn,
    author = "Sheykhi, Ahmad",
    title = "{Modified Friedmann Equations from Tsallis Entropy}",
    eprint = "1806.03996",
    archivePrefix = "arXiv",
    primaryClass = "gr-qc",
    doi = "10.1016/j.physletb.2018.08.036",
    journal = "Phys. Lett. B",
    volume = "785",
    pages = "118--126",
    year = "2018"
}

@article{Luciano:2023zrx,
    author = "Luciano, G. G.",
    title = "{From the emergence of cosmic space to horizon thermodynamics in Barrow entropy-based Cosmology}",
    doi = "10.1016/j.physletb.2023.137721",
    journal = "Phys. Lett. B",
    volume = "838",
    pages = "137721",
    year = "2023"
}

@article{P:2023jeq,
    author = "P, Nandhida Krishnan. and Mathew, Titus K.",
    title = "{Emergence of cosmic space with Barrow entropy, in non-equilibrium thermodynamic conditions}",
    eprint = "2302.01554",
    archivePrefix = "arXiv",
    primaryClass = "gr-qc",
    doi = "10.1016/j.dark.2023.101283",
    journal = "Phys. Dark Univ.",
    volume = "42",
    pages = "101283",
    year = "2023"
}

@article{Sheykhi:2023aqa,
    author = "Sheykhi, Ahmad",
    title = "{Corrections to Friedmann equations inspired by Kaniadakis entropy}",
    eprint = "2302.13012",
    archivePrefix = "arXiv",
    primaryClass = "gr-qc",
    doi = "10.1016/j.physletb.2024.138495",
    journal = "Phys. Lett. B",
    volume = "850",
    pages = "138495",
    year = "2024"
}

@article{Hernandez-Almada:2021rjs,
    author = "Hern{\'a}ndez-Almada, A. and Leon, Genly and Maga{\~n}a, Juan and Garc{\'\i}a-Aspeitia, Miguel A. and Motta, V. and Saridakis, Emmanuel N. and Yesmakhanova, Kuralay and Millano, Alfredo D.",
    title = "{Observational constraints and dynamical analysis of Kaniadakis horizon-entropy cosmology}",
    eprint = "2112.04615",
    archivePrefix = "arXiv",
    primaryClass = "astro-ph.CO",
    doi = "10.1093/mnras/stac795",
    journal = "Mon. Not. Roy. Astron. Soc.",
    volume = "512",
    number = "4",
    pages = "5122--5134",
    year = "2022"
}

@article{Hernandez-Almada:2021aiw,
    author = "Hern{\'a}ndez-Almada, A. and Leon, Genly and Maga{\~n}a, Juan and Garc{\'\i}a-Aspeitia, Miguel A. and Motta, V. and Saridakis, Emmanuel N. and Yesmakhanova, Kuralay",
    title = "{Kaniadakis-holographic dark energy: observational constraints and global dynamics}",
    eprint = "2111.00558",
    archivePrefix = "arXiv",
    primaryClass = "astro-ph.CO",
    doi = "10.1093/mnras/stac255",
    journal = "Mon. Not. Roy. Astron. Soc.",
    volume = "511",
    number = "3",
    pages = "4147--4158",
    year = "2022"
}

@article{Zarandi:2025ytt,
    author = "Zarandi, H. R. M. and Toda, Y. and Ebrahimi, E.",
    title = "{Investigating cosmology models through usual and dual kaniadakis entropies: theoretical and observational tensions and features}",
    doi = "10.1140/epjc/s10052-025-14307-6",
    journal = "Eur. Phys. J. C",
    volume = "85",
    number = "5",
    pages = "589",
    year = "2025"
}

@article{Odintsov:2024hzu,
    author = "Odintsov, Sergei D. and Paul, Tanmoy and SenGupta, Soumitra",
    title = "{Second law of horizon thermodynamics during cosmic evolution}",
    eprint = "2404.05312",
    archivePrefix = "arXiv",
    primaryClass = "gr-qc",
    doi = "10.1103/PhysRevD.109.103515",
    journal = "Phys. Rev. D",
    volume = "109",
    number = "10",
    pages = "103515",
    year = "2024"
}

@article{Asghari:2021bqa,
    author = "Asghari, Mahnaz and Sheykhi, Ahmad",
    title = "{Observational constraints of the modified cosmology through Barrow entropy}",
    eprint = "2110.00059",
    archivePrefix = "arXiv",
    primaryClass = "gr-qc",
    doi = "10.1140/epjc/s10052-022-10262-8",
    journal = "Eur. Phys. J. C",
    volume = "82",
    number = "5",
    pages = "388",
    year = "2022"
}

@article{Anagnostopoulos:2020ctz,
    author = "Anagnostopoulos, Fotios K. and Basilakos, Spyros and Saridakis, Emmanuel N.",
    title = "{Observational constraints on Barrow holographic dark energy}",
    eprint = "2005.10302",
    archivePrefix = "arXiv",
    primaryClass = "gr-qc",
    doi = "10.1140/epjc/s10052-020-8360-5",
    journal = "Eur. Phys. J. C",
    volume = "80",
    number = "9",
    pages = "826",
    year = "2020"
}

@article{Leon:2021wyx,
    author = "Leon, Genly and Maga{\~n}a, Juan and Hern{\'a}ndez-Almada, A. and Garc{\'\i}a-Aspeitia, Miguel A. and Verdugo, Tom{\'a}s and Motta, V.",
    title = "{Barrow Entropy Cosmology: an observational approach with a hint of stability analysis}",
    eprint = "2108.10998",
    archivePrefix = "arXiv",
    primaryClass = "astro-ph.CO",
    doi = "10.1088/1475-7516/2021/12/032",
    journal = "JCAP",
    volume = "12",
    number = "12",
    pages = "032",
    year = "2021"
}

@article{Lambiase:2023ryq,
    author = "Lambiase, Gaetano and Luciano, Giuseppe Gaetano and Sheykhi, Ahmad",
    title = "{Slow-roll inflation and growth of perturbations in Kaniadakis modification of Friedmann cosmology}",
    eprint = "2307.04027",
    archivePrefix = "arXiv",
    primaryClass = "gr-qc",
    doi = "10.1140/epjc/s10052-023-12112-7",
    journal = "Eur. Phys. J. C",
    volume = "83",
    number = "10",
    pages = "936",
    year = "2023"
}

@article{Jizba:2024klq,
    author = "Jizba, Petr and Lambiase, Gaetano and Luciano, Giuseppe Gaetano and Mastrototaro, Leonardo",
    title = "{Imprints of Barrow{\textendash}Tsallis cosmology in primordial gravitational waves}",
    eprint = "2403.09797",
    archivePrefix = "arXiv",
    primaryClass = "gr-qc",
    doi = "10.1140/epjc/s10052-024-13455-5",
    journal = "Eur. Phys. J. C",
    volume = "84",
    number = "10",
    pages = "1076",
    year = "2024"
}

@article{Padilla:2019mgi,
    author = "Padilla, Luis E. and Tellez, Luis O. and Escamilla, Luis A. and Vazquez, Jose Alberto",
    title = "{Cosmological Parameter Inference with Bayesian Statistics}",
    eprint = "1903.11127",
    archivePrefix = "arXiv",
    primaryClass = "astro-ph.CO",
    doi = "10.3390/universe7070213",
    journal = "Universe",
    volume = "7",
    number = "7",
    pages = "213",
    year = "2021"
}

@article{TANG2012383,
title = {Fractal dimension of pore-structure of porous metal materials made by stainless steel powder},
journal = {Powder Technology},
volume = {217},
pages = {383-387},
year = {2012},
issn = {0032-5910},
doi = {https://doi.org/10.1016/j.powtec.2011.10.053},
url = {https://www.sciencedirect.com/science/article/pii/S0032591011005961},
author = {H.P. Tang and J.Z. Wang and J.L. Zhu and Q.B. Ao and J.Y. Wang and B.J. Yang and Y.N. Li}
}

@article{article,
author = {Xu, Peng and Yu, Boming},
year = {2008},
month = {01},
pages = {74-81},
title = {Developing a new form of permeability and Kozeny–Carman constant for homogeneous porous media by means of fractal geometry},
volume = {31},
journal = {Advances in Water Resources},
doi = {10.1016/j.advwatres.2007.06.003}
}

@article{DAGOTTO1990268,
title = {Collapse of the wave function, anomalous dimensions and continuum limits in model scalar field theories},
journal = {Physics Letters B},
volume = {237},
number = {2},
pages = {268-273},
year = {1990},
issn = {0370-2693},
doi = {https://doi.org/10.1016/0370-2693(90)91442-E},
url = {https://www.sciencedirect.com/science/article/pii/037026939091442E},
author = {Elbio Dagotto and Aleksander Kocić and John B. Kogut}
}

@article{sharma1975new,
  title={New non-additive measures of entropy for discrete probability distributions},
  author={Sharma, Bhudev D and Mittal, Dharam P},
  journal={J. Math. Sci},
  volume={10},
  number={75},
  pages={28--40},
  year={1975}
}

@article{Ghaffari_2019,
   title={Black hole thermodynamics in Sharma–Mittal generalized entropy formalism},
   volume={51},
   ISSN={1572-9532},
   url={http://dx.doi.org/10.1007/s10714-019-2578-2},
   DOI={10.1007/s10714-019-2578-2},
   number={7},
   journal={General Relativity and Gravitation},
   publisher={Springer Science and Business Media LLC},
   author={Ghaffari, S. and Ziaie, A. H. and Moradpour, H. and Asghariyan, F. and Feleppa, F. and Tavayef, M.},
   year={2019},
   month=July }

@ARTICLE{AR308339,
	author = {Kolesnichenko, A. V.},
	title = {Entropic Cosmology Based on Kaniadakis Dual Entropy on the Cosmological Horizon of the Universe},
	journal = {Astronomy Reports},
        pages = {449--467},
	volume = {102},
	number = {6},
	year = {2025},

			issn = {0004-6299},

	url = {https://journals.rcsi.science/0004-6299/article/view/308339}
}

@article{Gohar:2023hnb,
    author = "Gohar, Hussain and Salzano, Vincenzo",
    title = "{On the foundations of entropic cosmologies: Inconsistencies, possible solutions and dead end signs}",
    eprint = "2307.01768",
    archivePrefix = "arXiv",
    primaryClass = "gr-qc",
    doi = "10.1016/j.physletb.2024.138781",
    journal = "Phys. Lett. B",
    volume = "855",
    pages = "138781",
    year = "2024"
}

@article{Davies:1987ti,
    author = "Davies, P. C. W.",
    title = "{Cosmological Horizons and the Generalized Second Law of Thermodynamics}",
    reportNumber = "NCL-87-TP7",
    doi = "10.1088/0264-9381/4/6/006",
    journal = "Class. Quant. Grav.",
    volume = "4",
    pages = "L225",
    year = "1987"
}

@article{Gohar:2023lta,
    author = "Gohar, Hussain and Salzano, Vincenzo",
    title = "{Generalized mass-to-horizon relation: A new global approach to entropic cosmologies and its connection to {\ensuremath{\Lambda}}CDM}",
    eprint = "2307.06239",
    archivePrefix = "arXiv",
    primaryClass = "gr-qc",
    doi = "10.1103/PhysRevD.109.084075",
    journal = "Phys. Rev. D",
    volume = "109",
    number = "8",
    pages = "084075",
    year = "2024"
}

@article{Prasanthan:2024fwg,
    author = "Prasanthan, Pranav and Nelleri, Sarath and Pradeepan, Archana K. and Poonthottathil, Navaneeth and Tom, Emmanuel",
    title = "{On the interrelation of the generalized holographic equipartition and entropy maximization in Kaniadakis paradigm}",
    eprint = "2410.04219",
    archivePrefix = "arXiv",
    primaryClass = "gr-qc",
    doi = "10.1007/s10714-024-03340-7",
    journal = "Gen. Rel. Grav.",
    volume = "57",
    number = "1",
    pages = "6",
    year = "2025"
}

@article{Krishna:2022zzw,
    author = "Krishna, P. B. and T., Hassan Basari V. and Mathew, Titus K.",
    title = "{Emergence of cosmic space and its connection with thermodynamic principles}",
    eprint = "2205.14868",
    archivePrefix = "arXiv",
    primaryClass = "gr-qc",
    doi = "10.1007/s10714-022-02941-4",
    journal = "Gen. Rel. Grav.",
    volume = "54",
    number = "6",
    pages = "58",
    year = "2022"
}

@article{T:2022mqr,
    author = "T., Hassan Basari V. and Krishna, P. B. and Mathew, Titus K.",
    title = "{Unified formalism for the law of emergence from the first law of thermodynamics}",
    eprint = "2209.00304",
    archivePrefix = "arXiv",
    primaryClass = "gr-qc",
    doi = "10.1103/PhysRevD.107.063511",
    journal = "Phys. Rev. D",
    volume = "107",
    number = "6",
    pages = "063511",
    year = "2023"
}

@article{Sardar:2024bfv,
    author = "Sardar, Alok and Debnath, Ujjal",
    title = "{Constraining the entropy corrected (m, n)-type pilgrim dark energy in fractal cosmology}",
    doi = "10.1140/epjc/s10052-024-13405-1",
    journal = "Eur. Phys. J. C",
    volume = "84",
    number = "10",
    pages = "1074",
    year = "2024"
}

@article{Carrasco:2023imi,
    author = "Carrasco, Raul and Rincon, Angel and Saavedra, Joel and Videla, Nelson",
    title = "{Discriminating interacting dark energy models using Statefinder diagnostic}",
    eprint = "2310.04324",
    archivePrefix = "arXiv",
    primaryClass = "gr-qc",
    doi = "10.1140/epjc/s10052-024-12733-6",
    journal = "Eur. Phys. J. C",
    volume = "84",
    number = "5",
    pages = "459",
    year = "2024"
}

@article{Bolotin:2026hot,
    author = "Bolotin, Yu. L. and Yanovsky, V., V. and Yerokhin, D. A.",
    title = "{Cosmographic connection between cosmological and Planck scales: the Barrow-Tsallis entropy}",
    eprint = "2602.12077",
    archivePrefix = "arXiv",
    primaryClass = "gr-qc",
    doi = "10.1088/1475-7516/2026/09/111",
    journal = "JCAP",
    volume = "09",
    pages = "111",
    year = "2026"
}

@article{Luciano:2026vhm,
    author = "Luciano, G. G. and Paliathanasis, A. and Leon, G. and Sheykhi, A. and Motaghi, M.",
    title = "{Hints beyond {\ensuremath{\Lambda}}CDM from Barrow and Tsallis holographic dark energy with GO cutoff}",
    eprint = "2601.02567",
    archivePrefix = "arXiv",
    primaryClass = "gr-qc",
    doi = "10.1016/j.physletb.2026.140154",
    journal = "Phys. Lett. B",
    volume = "873",
    pages = "140154",
    year = "2026"
}

\end{document}